\documentclass[sn-basic]{sn-jnl}% Math and Physical Sciences Reference Style
\jyear{2021}%

\usepackage{enumitem}
\usepackage[labelformat=empty, position=top]{subcaption}
\usepackage[export]{adjustbox}
\usepackage{tabularx}
\usepackage{pdflscape}
\usepackage{multirow}
\theoremstyle{thmstyleone}%
\theoremstyle{thmstyletwo}%

\theoremstyle{thmstylethree}%

\definecolor{cadmiumgreen}{rgb}{0.0, 0.42, 0.24}

\begin{document}

\title[~]{Leveraging Large Language Models (LLMs) for Process Mining (Technical Report)}

%%=============================================================%%
%% Prefix	-> \pfx{Dr}
%% GivenName	-> \fnm{Joergen W.}
%% Particle	-> \spfx{van der} -> surname prefix
%% FamilyName	-> \sur{Ploeg}
%% Suffix	-> \sfx{IV}
%% NatureName	-> \tanm{Poet Laureate} -> Title after name
%% Degrees	-> \dgr{MSc, PhD}
%% \author*[1,2]{\pfx{Dr} \fnm{Joergen W.} \spfx{van der} \sur{Ploeg} \sfx{IV} \tanm{Poet Laureate} 
%%                 \dgr{MSc, PhD}}\email{iauthor@gmail.com}
%%=============================================================%%

\author*[1,2]{\fnm{Alessandro} \sur{Berti}}\email{a.berti@pads.rwth-aachen.de}

%\author[2,1]{\fnm{Daniel} \sur{Schuster}}\email{daniel.schuster@fit.fraunhofer.de}

\author{\fnm{Mahnaz Sadat} \sur{Qafari}}\email{sadatghafari@gmail.com}

%\author[1,2]{\fnm{Wil M.P.} \sur{van der Aalst}}\email{wvdaalst@pads.rwth-aachen.de}

\affil*[1]{\orgdiv{Process and Data Science Group}, \orgname{RWTH Aachen University}, \orgaddress{\street{Ahornstrasse 55}, \city{Aachen}, \postcode{52074}, \state{NRW}, \country{Germany}}}
\affil*[2]{\orgdiv{Center for Process Intelligence}, \orgname{Fraunhofer FIT}, \orgaddress{\street{Schloss Birlinghoven, Konrad-Adenauer-Straße}, \city{Sankt Augustin}, \postcode{53757}, \state{NRW}, \country{Germany}}}

%In the field of data science, process mining plays a crucial role in extracting valuable insights from event data, ultimately improving business processes. The recent advancements in Large Language Models (LLMs), such as GPT-4, offer promising new avenues for the analysis and querying of these processes. However, the implementation of LLMs is challenged by their constraint in processing large datasets or comprehensive event logs, a limitation known as the 'context window'. To circumvent this challenge, we propose innovative methodologies that recast traditional process mining artifacts into digestible formats for LLMs. Our strategies focus on creating efficient prompts to exploit the significant potential of these models. We integrate our methodologies into the open-source process mining library, \emph{pm4py}, showcasing the versatility of our approach. Through various case studies using publicly available event logs, we demonstrate the effectiveness of our methods across diverse business processes such as medical, travel expense reporting, and fines management. This paper opens new doors in the intersection of process mining and natural language processing, highlighting the potential of LLMs in transforming our understanding and analysis of business processes.

\abstract{
This technical report describes the intersection of process mining and large language models (LLMs), specifically focusing on the abstraction of traditional and object-centric process mining artifacts into textual format. We introduce and explore various prompting strategies: direct answering, where the large language model directly addresses user queries; multi-prompt answering, which allows the model to incrementally build on the knowledge obtained through a series of prompts; and the generation of database queries, facilitating the validation of hypotheses against the original event log.

Our assessment considers two large language models, GPT-4 and Google's Bard, under various contextual scenarios across all prompting strategies. Results indicate that these models exhibit a robust understanding of key process mining abstractions, with notable proficiency in interpreting both declarative and procedural process models.

In addition, we find that both models demonstrate strong performance in the object-centric setting, which could significantly propel the advancement of the object-centric process mining discipline.

Additionally, these models display a noteworthy capacity to evaluate various concepts of fairness in process mining. This opens the door to more rapid and efficient assessments of the fairness of process mining event logs, which has significant implications for the field.

The integration of these large language models into process mining applications may open new avenues for exploration, innovation, and insight generation in the field.
}

\keywords{Large Language Models, Process Mining, GPT-4, Querying Language}

\maketitle

\lstset{
  basicstyle=\scriptsize,
  columns=fullflexible,
  frame=single,
  breaklines=true,
  %postbreak=\mbox{\textcolor{red}{$\hookrightarrow$}\space},
}

\iffalse

\section{Introduction}
\label{sec:introduction}

\fi

\section{Preliminaries}
\label{sec:preliminaries}

The following subsections delve into some key preliminaries to provide the necessary background for our discussion. This encompasses the nature of traditional and object-centric event logs (\autoref{subsec:tradObjCentrEvLogs}), the distinctions between procedural and declarative process models (\autoref{subsec:procedProcessModels}), the techniques for extracting numerical features from event logs (\autoref{subsec:feaExt}), and the understanding of fairness concepts within the scope of process mining (\autoref{subsec:fairnessConcepts}). These topics form the cornerstone of our exploration into the integration of large language models within process mining.

\subsection{Traditional and Object-Centric Event Logs}
\label{subsec:tradObjCentrEvLogs}

\begin{table}[h!]
\centering
\begin{tabular}{|c|c|c|}
\hline
\textbf{Case ID} & \textbf{Activity} & \textbf{Timestamp} \\
\hline
Case1 & Activity A & 2023-07-13 09:00:00 \\
Case1 & Activity B & 2023-07-13 09:15:00 \\
Case1 & Activity C & 2023-07-13 09:30:00 \\
Case2 & Activity A & 2023-07-13 10:00:00 \\
Case2 & Activity B & 2023-07-13 10:30:00 \\
Case3 & Activity A & 2023-07-13 11:00:00 \\
Case3 & Activity C & 2023-07-13 11:30:00 \\
Case3 & Activity D & 2023-07-13 11:45:00 \\
\hline
\end{tabular}
\caption{Simple example of a traditional event log}
\label{table:traditionalEventLog}
\end{table}

Event logs are pivotal to process mining, representing empirical data that drives operations such as process discovery, conformance checking, and model enhancement. Extracted from information systems, these logs comprise collections of events, each characterized by an activity and timestamp and associated with a particular process. As primary artifacts, event logs provide a robust basis for data-driven process analysis.

Traditional event logs adhere to a format where each event is tied to a specific case, signifying a unique execution of the process. Every event is defined by its activity—representing the operation performed, and a timestamp—indicating when the operation took place. The Extensible Event Stream (XES) standard \cite{DBLP:journals/cim/AcamporaVSAGV17} has been established for storing and exchanging these event logs, facilitating systematic documentation and analysis of process events.
An example of traditional event logs is reported in \autoref{table:traditionalEventLog}. We have three cases. For \emph{Case1}, we have three events with activities A,B,C.
For \emph{Case2}, we have two events with activities A and B. For \emph{Case3}, we have three events with activities A,C,D.

\begin{table}[h!]
\centering
\begin{tabular}{|c|c|c|c|c|c|}
\hline
\textbf{Event ID} & \textbf{Activity} & \textbf{Timestamp} & \textbf{Order} & \textbf{OrderItems} & \textbf{Invoices} \\
\hline
Event1 & Activity A & 2023-07-13 09:00:00 & {Order1} & {Item1, Item2} & {Invoice1} \\
Event2 & Activity B & 2023-07-13 09:15:00 & {Order1} & {Item1, Item2} & {Invoice1} \\
Event3 & Activity C & 2023-07-13 09:30:00 & {Order2} & {Item3} & {Invoice2} \\
Event4 & Activity A & 2023-07-13 10:00:00 & {Order2} & {Item3} & {Invoice2} \\
Event5 & Activity B & 2023-07-13 10:30:00 & {Order2} & {Item4} & {Invoice2} \\
Event6 & Activity A & 2023-07-13 11:00:00 & {Order3} & {Item5, Item6} & {Invoice3} \\
Event7 & Activity C & 2023-07-13 11:30:00 & {Order3} & {Item5} & {Invoice3} \\
Event8 & Activity D & 2023-07-13 11:45:00 & {Order3} & {Item6} & {Invoice3} \\
Event9 & Activity A & 2023-07-13 12:00:00 & {Order4} & {Item7, Item8} & {Invoice4} \\
Event10 & Activity B & 2023-07-13 12:15:00 & {Order4} & {Item8} & {Invoice4} \\
\hline
\end{tabular}
\caption{Simple example of an object-centric event log}
\label{table:objectCentricEventLog}
\end{table}

Object-centric event logs, on the other hand, cater to more complex scenarios by associating each event with multiple objects of different types. Each event still holds information about its activity and timestamp, but instead of relating to a single process execution, it ties into multiple entities. The OCEL standard \cite{DBLP:conf/adbis/GhahfarokhiPBA21} has been proposed as a standard for the storage of object-centric event logs, capturing detailed interaction patterns between diverse process entities.
An example is reported in \autoref{table:objectCentricEventLog}. For example, we see that the event with identifier \emph{Event1} is related to one object of type order, two objects of type order item, and one object of type invoice.

Object-centric event logs allow for the capture of complex, intertwined, multi-object scenarios, opening avenues for more nuanced process mining tasks. However, the complexity of these logs presents unique challenges for process mining, such as how to structure the data for analysis. A significant stride has been made in addressing these challenges with the proposal by Adams et al. in their 2022 study \cite{DBLP:conf/icpm/AdamsSSSA22} to partition object-centric event logs into "executions".
Their methodology, inspired by traditional case-oriented process mining, breaks down the intricate graph structure of object-centric event logs into manageable "process executions", effectively the equivalent of cases in traditional event logs. By extending the case concept to object-centric data, the authors enable a more refined application of process mining tasks to these logs, enabling practitioners to query specific aspects of a process. Furthermore, through the application of graph isomorphism, they are able to identify equivalent process executions with respect to an attribute, thereby discovering object-centric variants.
This approach offers a method for handling the complexities inherent in object-centric event logs without resorting to flattening the data, thereby preserving information and structure.
As example, the object-centric event log in \autoref{table:objectCentricEventLog} is split in four distinct connected components (
$\{ \textit{Order1}, \textit{Item1}, \textit{Item2}, \textit{Invoice1} \}$,
$\{ \textit{Order2}, \textit{Item3}, \textit{Invoice2} \}$,
$\{ \textit{Order3}, \textit{Item5}, \textit{Item6}, \textit{Invoice3} \}$,
$\{ \textit{Order4}, \textit{Item8}, \textit{Invoice4} \}$ ).

\begin{figure}[ht]
\centering
\begin{subfigure}{0.45\textwidth}
  \centering
  \includegraphics[height=65mm]{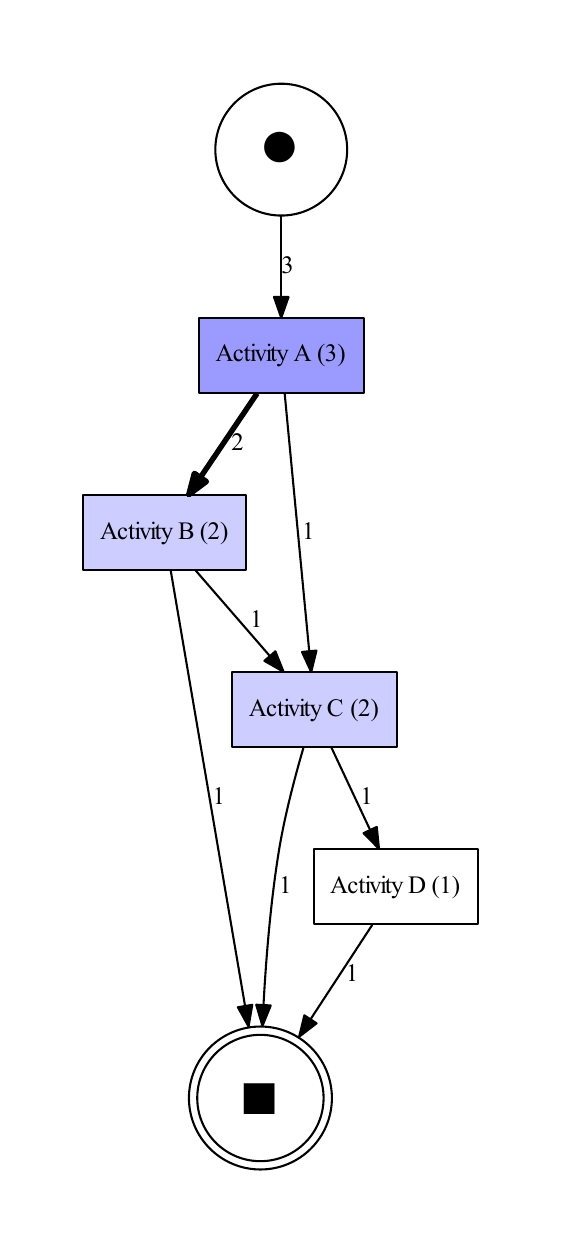}
  \caption{Frequency Annotation}
  \label{fig:exampleDfgFrequency}
\end{subfigure}%
\hfill
\begin{subfigure}{0.45\textwidth}
  \centering
  \includegraphics[height=65mm]{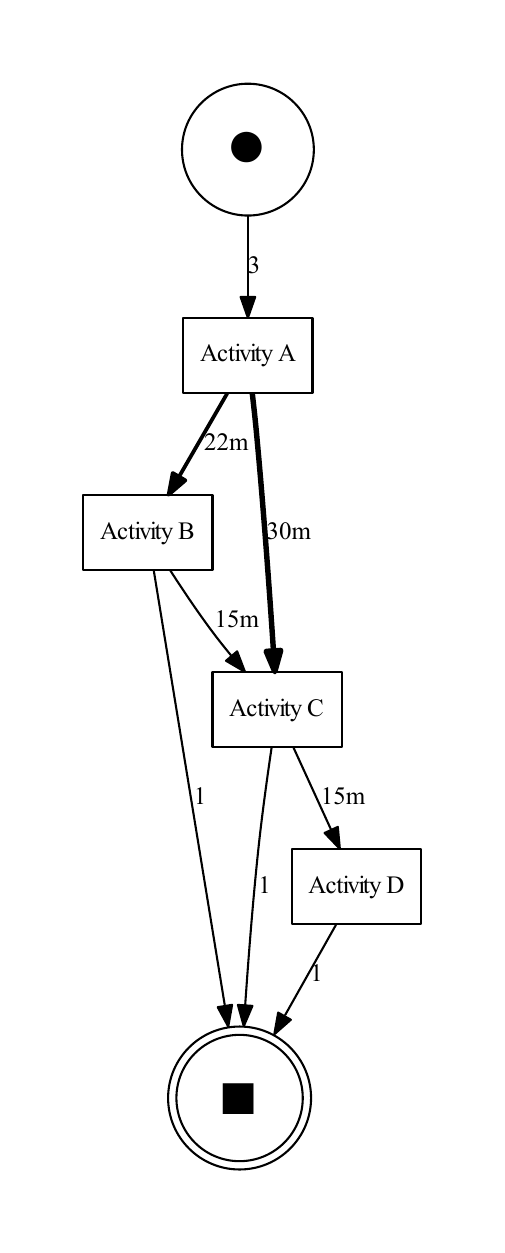}
  \caption{Performance Annotation}
  \label{fig:exampleDfgPerformance}
\end{subfigure}
\caption{Example directly-follows graph (DFG) computed on the event log described in \autoref{table:traditionalEventLog}.}
\label{fig:exampleDfg}
\end{figure}

\subsection{Procedural Process Models}
\label{subsec:procedProcessModels}

Procedural process models, as a fundamental tool in process mining, systematically depict the order of activities and their interrelations within a process. These models capture the idea that some activities must precede others, clearly indicating the sequence or procedure that needs to be followed to attain a particular outcome. They represent a step-by-step visualization of a process, guiding us from the beginning to the end, hence the name 'procedural.' In such models, each step is contingent on the output of the preceding one, reinforcing a precise flow of tasks in a process. Now, let's delve into some common types of procedural process models.

\begin{figure}[ht]
\centering
\includegraphics[width=\textwidth]{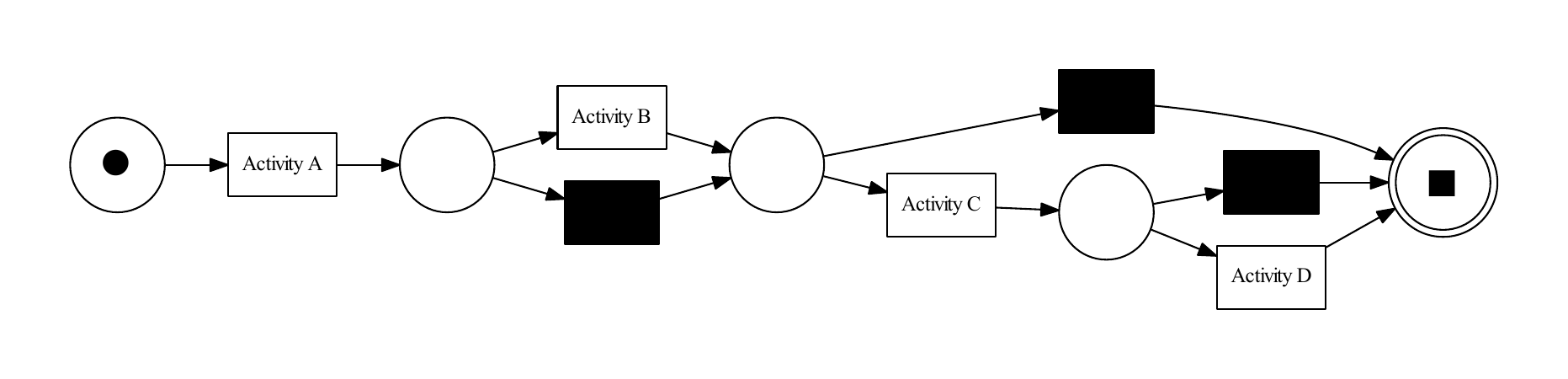}
\caption{Example Petri net obtained from the event log described in \autoref{table:traditionalEventLog} using the Inductive miner process discovery algorithm \cite{inductiveminer}.}
\label{fig:examplePetriNet}
\end{figure}

A primary example of a procedural process model is the directly-follows graph (DFG) \cite{DBLP:books/sp/Aalst16}. This model encapsulates the succession relationship between activities, where an edge from activity A to activity B signifies that B has directly followed A in the process. DFGs provide an intuitive, visual way to comprehend the sequence of activities in a process and can be an effective starting point for process discovery.
An example of a directly-follows graph, annotated with frequency and performance information, is reported in \autoref{fig:exampleDfg}.

\begin{figure}[ht]
\centering
\includegraphics[width=\textwidth]{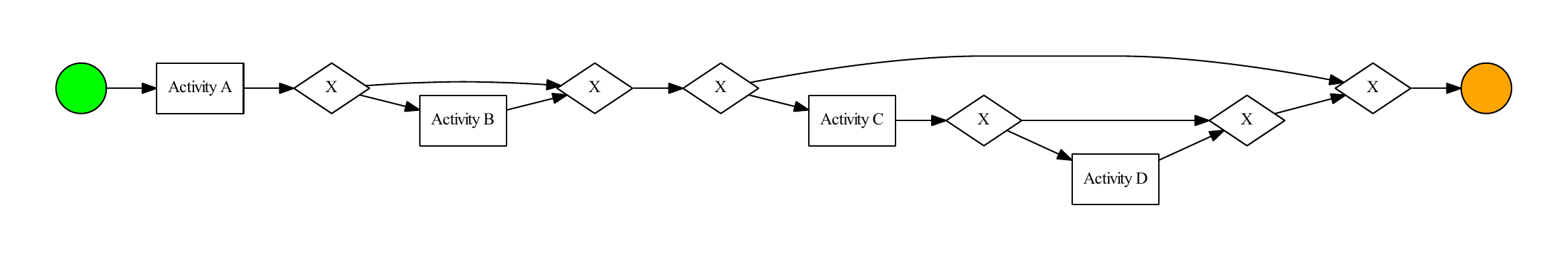}
\caption{Example BPMN model obtained from the event log described in \autoref{table:traditionalEventLog} using the Inductive miner process discovery algorithm \cite{inductiveminer}.}
\label{fig:exampleBPMN}
\end{figure}

In the realm of formal procedural process models, Petri nets have gained wide acceptance in process mining \cite{DBLP:journals/jcsc/Aalst98}. They offer a graphical and mathematical modelling tool well suited to capture concurrency, synchronization and shared resources. In a Petri net, places (represented as circles) can contain tokens, and transitions (represented as rectangles) can consume and produce these tokens, in line with predefined rules.

Different process discovery algorithms have Petri nets as eventual output, including the Alpha \cite{alpha}, the Inductive \cite{inductiveminer} and the Heuristics \cite{heuristicsminer} miners.
An example of Petri net model, discovered from \autoref{table:traditionalEventLog} using the Inductive miner, is presented in \autoref{fig:examplePetriNet}.

Another common procedural modelling notation used in business process management is the Business Process Model and Notation (BPMN 2.0 \cite{DBLP:series/ihis/AagesenK15}). It is a rich, graphical representation for specifying business processes in a business process model, offering a more extensive set of symbols compared to Petri nets, thereby allowing more precise business communication \cite{DBLP:journals/sosym/KalenkovaALR17,DBLP:journals/bpmj/KalenkovaBLAS19}.
However, the expressiveness of BPMN can lead to ambiguity due to the potential for different interpretations of its constructs \cite{DBLP:journals/infsof/DijkmanDO08,DBLP:conf/icws/OuyangDHA06}.
An example of BPMN model is represented in \autoref{fig:exampleBPMN}. We also provide an example of BPMN 2.0 XML\footnote{Available at the address \url{https://github.com/pm4py/pm4py-core/blob/release/tests/input_data/running-example.bpmn}}.

\begin{figure}[ht]
\centering
\begin{subfigure}{0.7\textwidth}
  \centering
  \includegraphics[width=\linewidth]{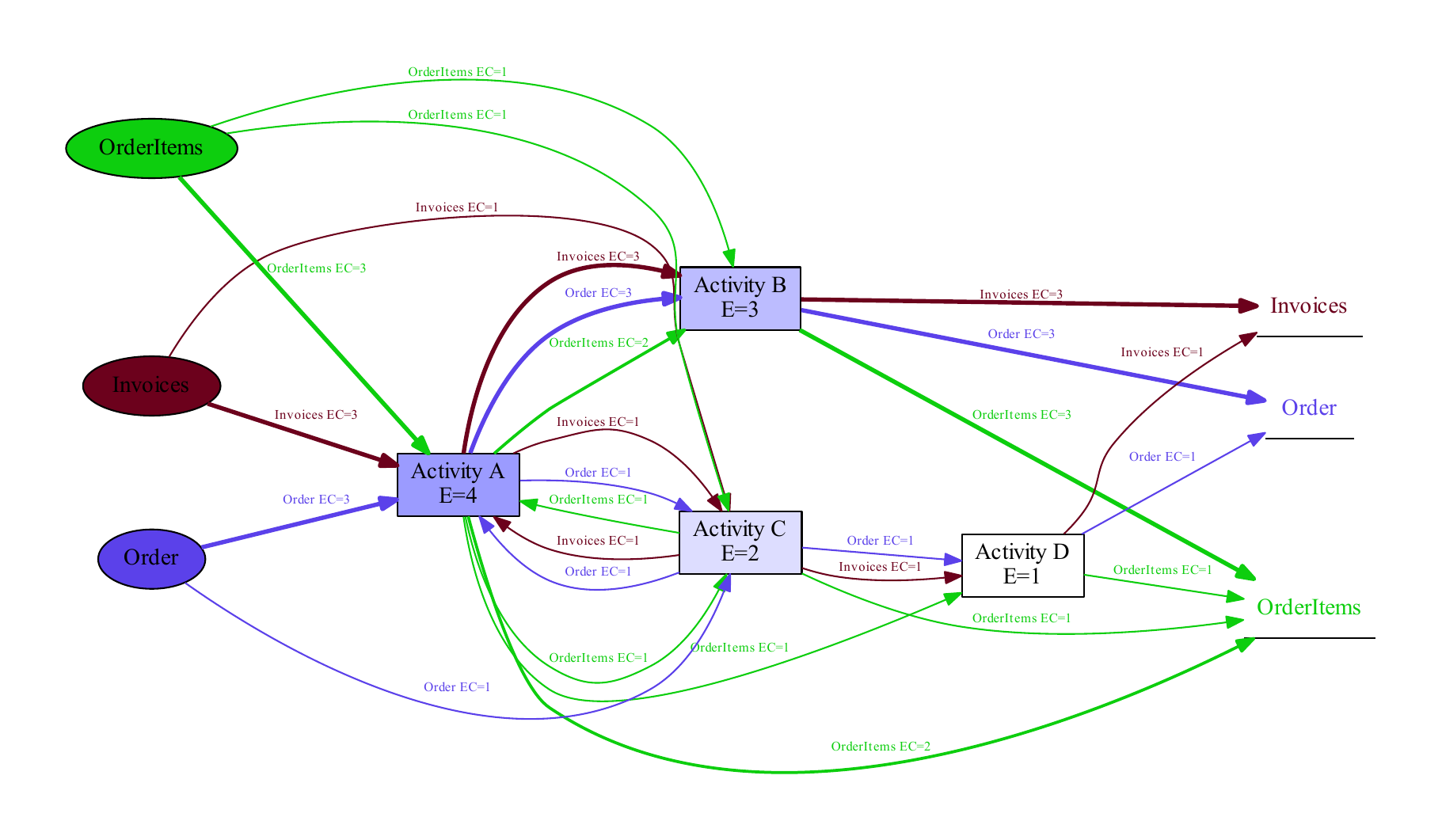}
  \caption{Frequency Annotation}
  \label{fig:exampleOCDFG1}
\end{subfigure}

\begin{subfigure}{0.7\textwidth}
  \centering
  \includegraphics[width=\linewidth]{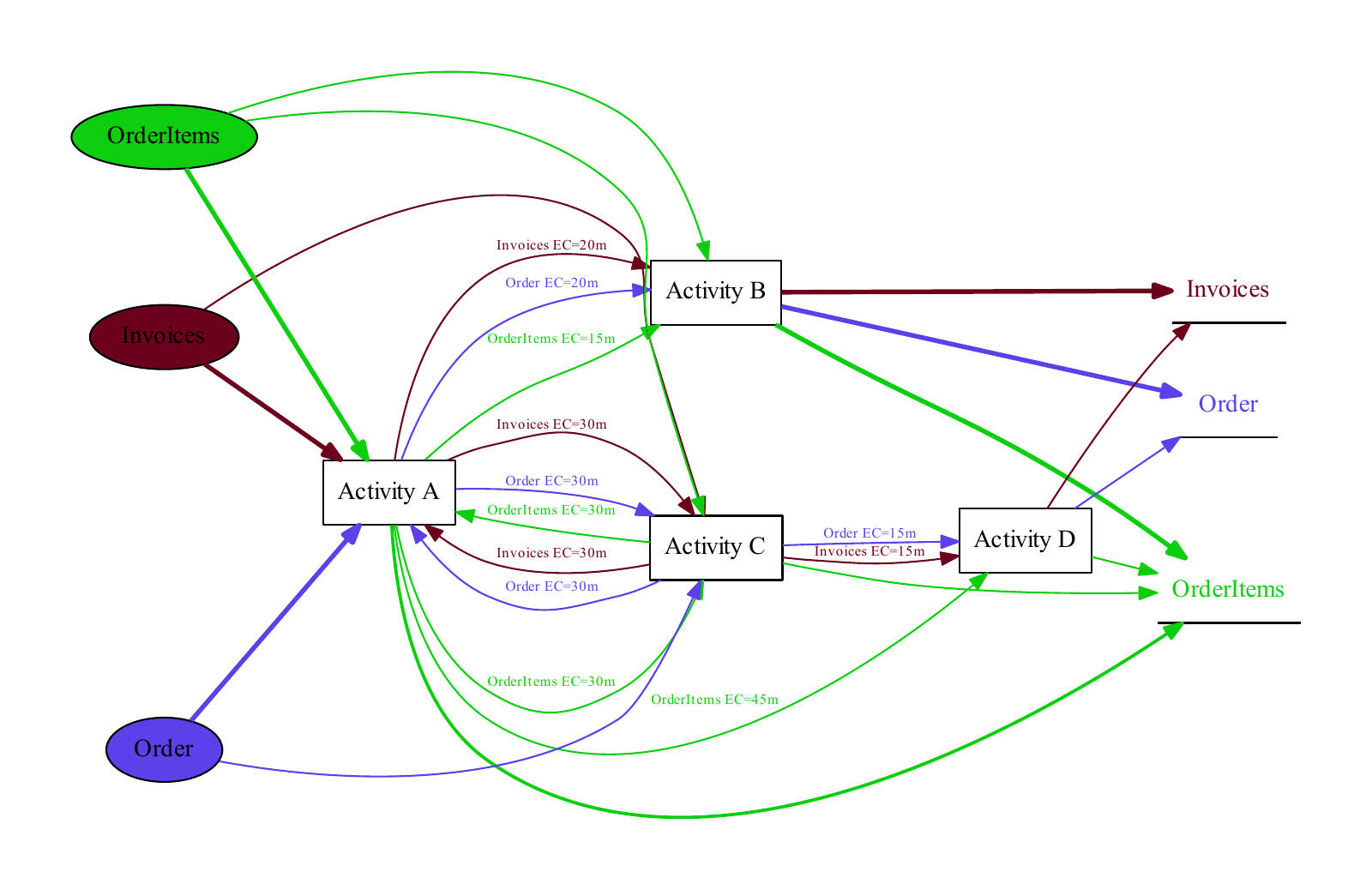}
  \caption{Performance Annotation}
  \label{fig:exampleOCDFG2}
\end{subfigure}
\caption{Example object-centric directly-follows graph computed on the object-centric log contained in \autoref{table:objectCentricEventLog}. Different annotations (i.e., frequency and performance) are computed \cite{DBLP:journals/sttt/BertiA23}.}
\label{fig:exampleOCDFG}
\end{figure}

In the realm of object-centric process mining, procedural process models extend their scope to encapsulate the multiple object types involved in a process. This leads to the creation of two key variants: object-centric directly-follows graphs and object-centric Petri nets.
An \emph{object-centric directly-follows graph} \cite{DBLP:journals/sttt/BertiA23} represents a direct evolution from the traditional directly-follows graph, with each object type having its own graph portraying the succession of activities involving that particular object. This collation of graphs offers a nuanced view of the process, maintaining the simplicity of the traditional format.
On the other hand, the \emph{object-centric Petri net} \cite{DBLP:journals/fuin/AalstB20} extends traditional Petri nets to capture the lifecycle of each individual object type within the process. Each object type is represented by its own Petri net, and their integration forms the object-centric Petri net. This detailed representation provides a comprehensive view of the process, accounting for the interactions between the multiple object types.
These models provide a powerful means for analyzing complex, object-centric processes, enabling an in-depth understanding of the process from the perspective of individual object types and their interaction.

\subsection{Declarative Process Models}
\label{subsec:declarativeProcModels}

Declarative process models provide a different perspective compared to the traditional procedural process models. While procedural models specify the explicit sequence of activities in a process, declarative models describe the process through a set of constraints or rules that prescribe what cannot happen, leaving a degree of flexibility for what can happen.

A common notation used for declarative modeling in the process mining field is the DECLARE framework \cite{DBLP:conf/cidm/MaggiMA11}. It targets the control-flow perspective of a process by specifying constraints on the sequence, choice, and parallelism of activities. For instance, a DECLARE model can express that if activity 'A' happens, then activity 'B' must eventually follow, capturing the interdependence between activities.

\begin{table}[h!]
\centering
\begin{tabular}{|c|c|c|c|}
\hline
\textbf{Source Activity} & \textbf{Target Activity} & \textbf{Average Time} & \textbf{Standard Deviation} \\
\hline
Activity A & Activity B & 1350.0 & 636.40 \\
Activity A & Activity C & 1800.0 & 0.00 \\
Activity B & Activity C & 900.0 & 0.00 \\
Activity A & Activity D & 2700.0 & 0.00 \\
Activity C & Activity D & 900.0 & 0.00 \\
\hline
\end{tabular}
\caption{Temporal profile model \cite{DBLP:journals/corr/abs-2008-07262} for the traditional event log presented in \autoref{table:traditionalEventLog}.}
\label{table:temporalProfile}
\end{table}

Furthermore, declarative process models extend beyond control-flow constraints to capture other aspects of the process. Temporal declarative models incorporate temporal constraints into the process \cite{DBLP:journals/corr/abs-2008-07262,DBLP:conf/caise/BarrientosWMR23}, providing further refinement on when activities should occur. Data-driven declarative models integrate data conditions to determine the valid sequence of activities, thus introducing a context-sensitive dimension to the process model.
An example temporal model is presented in \autoref{table:temporalProfile}.

Overall, declarative process models offer a flexible and holistic approach to process modeling, allowing for a wide range of scenarios and conditions to be incorporated, thereby addressing the complex, dynamic nature of many real-world processes.

\subsection{Extraction of Numerical Features}
\label{subsec:feaExt}

Feature extraction is a fundamental process in transforming raw event log data into a format that machine learning algorithms can interpret and learn from. In the context of process mining, feature extraction plays a significant role in various tasks, including predictive process monitoring, anomaly detection, and process enhancement, among others \cite{DBLP:journals/is/LeoniAD16}.

The primary strategy to convert an event log into numeric features is through the use of encoding techniques such as one-hot encoding. In this strategy, each unique activity in the event log is assigned a distinct binary (0 or 1) feature, thus creating a binary vector representation for each case in the log. This enables machine learning algorithms to effectively capture and learn the presence or absence of specific activities within each case. This transformation is crucial since machine learning algorithms primarily work with numeric data.
A numeric feature table computed on the traditional event log in \autoref{table:traditionalEventLog} is presented in \autoref{table:numericFeatureTable}.

Apart from the straightforward one-hot encoding, more advanced feature extraction methods can be considered. For instance, there are aggregate features that summarize specific characteristics of a case, like the total number of events, the mean time between events, or the standard deviation of the time between events. Another advanced technique is sequence encoding, where the order of activities is preserved. This is particularly beneficial when the sequence of activities is of paramount importance to the process. Other techniques include using n-grams \cite{DBLP:conf/icpm/ZandkarimiRSH20}, which capture sequences of 'n' consecutive activities, or using methods from natural language processing, such as word embeddings \cite{DBLP:conf/cikm/JuniorCDT20}, to encode activities and their context within the case.

\begin{table}[h!]
\centering
\resizebox{\textwidth}{!}{%
\begin{tabular}{|c|c|c|c|c|c|c|}
\hline
\textbf{Case ID} & \textbf{Activity A} & \textbf{Activity B} & \textbf{Activity C} & \textbf{Activity D} & \textbf{Count} & \textbf{Total Duration} \\
\hline
Case1 & 1 & 1 & 1 & 0 & 3 & 30 \\
Case2 & 1 & 1 & 0 & 0 & 2 & 30 \\
Case3 & 1 & 0 & 1 & 1 & 3 & 45 \\
\hline
\end{tabular}%
}
\caption{Numeric feature table for the traditional event log presented in \autoref{table:traditionalEventLog}.}
\label{table:numericFeatureTable}
\end{table}

In the realm of object-centric event logs, numeric feature extraction expands to accommodate the complexity and multidimensionality of these logs. Features are designed to represent the intricate dynamics involving multiple object types, their interactions, and temporal associations \cite{berti2022graph,DBLP:conf/icsoc/AdamsPLSA22}.
A numeric feature table computed on the object-centric event log in \autoref{table:objectCentricEventLog} is presented in \autoref{table:numericFeatureTableObjectCentric}.

Activity features from object-centric event logs encapsulate the actions occurring on specific objects or types of objects. For each unique activity, a binary indicator or count variable can be created to signify whether a particular activity has occurred or how often it has taken place on each object. This not only helps encapsulate the inherent processes but also provides a granular understanding of different object types' behaviors.

Inter-object features delve into the relationships between different objects in an object-centric log. When objects of different types consistently appear together, a feature is devised to symbolize this co-occurrence. This might be a binary flag signaling the presence of such a relationship or a count of the frequency of these interactions, offering insight into patterns of object interdependencies and interplay.

Temporal features are instrumental in encapsulating the timing aspects associated with events. For each object type, these can include the timestamp of the first or last event, the mean time between events, or the overall duration of processes involving the object. The temporal features help shed light on the pace and sequence of events across different objects, an essential aspect of understanding the overall process flow.

\begin{table}[h!]
\centering
\resizebox{\textwidth}{!}{%
\begin{tabular}{|c|c|c|c|c|c|c|}
\hline
\textbf{Order} & \textbf{Activity A} & \textbf{Activity B} & \textbf{Activity C} & \textbf{Activity D} & \textbf{Related Items} & \textbf{Related Invoices} \\
\hline
Order1 & 1 & 1 & 0 & 0 & 2 & 1 \\
Order2 & 1 & 1 & 1 & 0 & 2 & 1 \\
Order3 & 1 & 0 & 1 & 1 & 2 & 1 \\
Order4 & 1 & 1 & 0 & 0 & 2 & 1 \\
\hline
\end{tabular}}
\caption{Numeric feature table (for the object type \emph{Order}) derived from the object-centric event log presented in \autoref{table:objectCentricEventLog}.}
\label{table:numericFeatureTableObjectCentric}
\end{table}

Finally, data attribute features cater to any additional attributes tied to the events or objects themselves. If an event or object comes with associated data attributes such as cost or resource allocation, these attributes can be incorporated as features. This inclusion ensures any vital context-specific or process-related information is not overlooked during the feature extraction process \cite{DBLP:conf/caise/QafariA21}, thereby enhancing the explanatory power of the derived features.

By generating suitable features from event logs, we can create informative inputs for machine learning algorithms, enhancing their ability to model and predict complex process behaviors.

\subsection{Fairness Concepts in Process Mining}
\label{subsec:fairnessConcepts}

Fairness in process mining is a central theme of increasing importance \cite{DBLP:conf/otm/QafariA19}. In essence, it aims to ensure that the algorithms and models developed based on event data do not unfairly favor or disadvantage certain instances or groups. This is crucial because process mining techniques are often used to inform operational decisions and business strategies that can significantly impact people and resources. For example, they may be used to optimize workflows, allocate resources, or make predictive decisions. If these models or algorithms are biased or unfair, the resulting decisions can be unjust, leading to significant negative consequences. Therefore, maintaining fairness in process mining is not just an ethical imperative, but also a necessary condition for building trust and ensuring the overall sustainability of business processes.
Fairness in process mining can be thought of as a continuum that spans from the individual to the group level, incorporating notions of procedural and counterfactual fairness along the way \cite{DBLP:conf/icpm/PohlQA22}.

At the \emph{individual level}, fairness centers on ensuring that every case in a process is treated equitably. This could involve examining each instance independently to determine whether the same rules and decisions were applied consistently. For example, in a loan application process, individual fairness would imply that each application is evaluated based on the same set of criteria and not influenced by irrelevant attributes such as the applicant's race or gender.

Progressing towards the \emph{group level}, fairness becomes about ensuring equitable outcomes across different groups within a process. This could mean analyzing whether certain groups are systematically disadvantaged compared to others. For example, in an employee onboarding process, group fairness would imply that new hires from all backgrounds receive equal access to resources and opportunities for training and development.

The notion of \emph{procedural fairness} emphasizes the importance of the process itself being fair, beyond merely the outcomes it produces. Procedural fairness is concerned with the transparency, consistency, and unbiased nature of the processes. For instance, in a complaint handling process, procedural fairness would mean that all complaints are handled following the same, predefined steps and procedures without any bias.

\emph{Counterfactual fairness} \cite{DBLP:conf/caise/QafariA21} introduces a more complex, hypothetical element into the fairness equation. It considers whether the outcome of a process would have been the same had an irrelevant attribute been different. For example, in an employee promotion process, counterfactual fairness would mean that an employee's promotion decision would have been the same, irrespective of their gender.

One of the key challenges in the development of fairness-oriented techniques in process mining is the scarcity of publicly available event logs that feature fairness concerns. Such data sets are crucial for testing and refining techniques. They not only provide a ground truth against which the effectiveness of methods can be evaluated, but also highlight the real-world complexities and nuances that such techniques need to handle. Currently, the only public collection of simulated event logs that are designed to exhibit fairness issues is provided in \cite{pohl2023collection}. These logs and their characteristics are described in detail in the paper \cite{DBLP:journals/corr/abs-2306-11453}. This availability gap underlines the need for greater openness and collaboration in the field, and the importance of creating and sharing more such resources, so as to advance the development and validation of fairness-oriented process mining techniques.

\section{Approach}
\label{sec:approach}

In the forthcoming section, we unpack our unique approach to leveraging large language models for process mining tasks. Initially, we focus on developing textual abstractions for process mining artifacts \autoref{subsec:textAbstrProMinArtfs}, a vital foundation for enabling the language model's comprehension of process-related entities. This step includes particular emphasis on process models and event logs \autoref{subsubsec:textAbstrProMinModels} \autoref{subsubsec:textAbstrProMinEvLogs}. Subsequently, we delineate various prompting strategies \autoref{subsec:promptingStrategies} to guide the model in problem resolution. These strategies span from direct answering to multi-prompt answering, and even generation of database queries \autoref{subsubsec:strategyDirectAnswering} \autoref{subsubsec:strategyMultiPromptAnswering} \autoref{subsubsec:strategyGenerationDbQueries}. Through an effective synergy between these components, we navigate the nuanced landscape of process mining with large language models, ensuring efficient analysis and meaningful insights.

\subsection{Textual Abstractions of Process Mining Artifacts}
\label{subsec:textAbstrProMinArtfs}

In order to leverage the capabilities of Large Language Models (LLMs) for process mining, we need to convert process mining artifacts into a format that these models can understand and interpret effectively. The initial and crucial step in this transformation process is the textual encoding of these artifacts. The goal of this encoding is to convey the essential information embedded in the artifacts while maintaining the constraints imposed by the context window of the LLM.

\begin{figure}[H]
\lstset{caption={Textual abstraction of the directly-follows graph presented in \autoref{fig:exampleDfg} (obtained using the method \texttt{pm4py.llm.abstract\_dfg}).}}
\lstset{label={lst:dfgTextualAbstraction}}
\begin{lstlisting}
If I have a process with flow:

Activity A -> Activity B ( frequency = 2  performance = 1350.0 )
Activity C -> Activity D ( frequency = 1  performance = 900.0 )
Activity B -> Activity C ( frequency = 1  performance = 900.0 )
Activity A -> Activity C ( frequency = 1  performance = 1800.0 )
\end{lstlisting}
\end{figure}

In our research, we have developed distinct textual abstractions for different types of event logs and process models. On the one hand, we address traditional and object-centric event logs, which provide a chronological record of events related to a particular process. On the other hand, we also consider procedural and declarative process models. Procedural models focus on the sequence of activities necessary to accomplish a process, while declarative models describe the rules and constraints that govern the execution of the process. Our textual abstractions offer a balanced summary of these artifacts that preserve their critical properties while remaining accessible to LLMs.

\subsubsection{Textual Abstractions of Process Models}
\label{subsubsec:textAbstrProMinModels}

One of the fundamental process mining artifacts is the directly-follows graph (DFG) introduced in \autoref{subsec:procedProcessModels}. To translate this information into a form digestible by Large Language Models (LLMs), we need to convert the DFG into a textual abstraction. This conversion, as exemplified in \autoref{lst:dfgTextualAbstraction}, is a careful process that aims to maintain the core information contained in the graph while conforming to the text-based input restrictions of the LLMs.
Each arc in the DFG is translated into a textual sentence that describes the transition from one activity to another, along with its frequency and performance. Frequency signifies how often a particular sequence of activities occurs in the process, while performance captures the average time it takes to move from one activity to the next.
The arcs are listed in order of decreasing frequency to prioritize the most common transitions in the process. If the total length of the textual abstraction threatens to exceed the context window limitation of the LLM, some of the less frequent arcs may be omitted. This ensures that the model is fed with the most representative information about the process within the imposed text length constraint, thus enabling effective query answering and process analysis.
In addition to the standard textual encoding of the DFG, it is worth noting that advanced large language models exhibit significant flexibility in understanding various text representations of process flows. They are capable of interpreting and reasoning about different encodings for the same underlying information. For instance, the transition from "Activity A" to "Activity B" could be expressed in several ways, such as "Activity A is followed by Activity B", "Activity A $\rightarrow$ Activity B", or even as a pair "(Activity A, Activity B)". Despite the stylistic differences, the LLMs can recognize these expressions as conveying the same transition in the process. This flexibility makes LLMs robust and adaptable tools for analyzing process mining artifacts represented in different textual formats.

\begin{figure}[H]
\lstset{caption={Textual abstraction of the Petri net presented in \autoref{fig:examplePetriNet} (obtained using the method \texttt{pm4py.llm.abstract\_petri\_net}).}}
\lstset{label={lst:petriNetTextualAbstraction}}
\begin{lstlisting}
If I have a Petri net:

places: [ p_3, p_4, p_5, sink, source ]
transitions: [ (0309dbf9-52f9-42a5-a73a-79da3a2ea20a, 'Activity B'), (37ede306-4d81-499a-8dbe-b23b5a31372a, 'Activity C'), (5f5a9dd2-b140-4e88-ac73-b05c3ab17b63, 'Activity A'), (eba11fc0-143b-4765-adb0-2b736b348f88, 'Activity D'), (skip_1, None), (skip_2, None), (skip_3, None) ]
arcs: [ (0309dbf9-52f9-42a5-a73a-79da3a2ea20a, 'Activity B')->p_4, (37ede306-4d81-499a-8dbe-b23b5a31372a, 'Activity C')->p_5, (5f5a9dd2-b140-4e88-ac73-b05c3ab17b63, 'Activity A')->p_3, (eba11fc0-143b-4765-adb0-2b736b348f88, 'Activity D')->sink, (skip_1, None)->p_4, (skip_2, None)->sink, (skip_3, None)->sink, p_3->(0309dbf9-52f9-42a5-a73a-79da3a2ea20a, 'Activity B'), p_3->(skip_1, None), p_4->(37ede306-4d81-499a-8dbe-b23b5a31372a, 'Activity C'), p_4->(skip_2, None), p_5->(eba11fc0-143b-4765-adb0-2b736b348f88, 'Activity D'), p_5->(skip_3, None), source->(5f5a9dd2-b140-4e88-ac73-b05c3ab17b63, 'Activity A') ]

initial marking: ['source:1']
final marking: ['sink:1']
\end{lstlisting}
\end{figure}

In our process of creating a textual representation of Petri nets (\autoref{subsec:procedProcessModels}), we assign unique identifiers to both places and transitions within the net, as demonstrated in \autoref{lst:petriNetTextualAbstraction}. Visible transitions are accompanied by a specific label, while arcs serve as the connectors between places and transitions and are depicted accurately. This approach enables a clear and precise representation of the Petri net's structure, making it easy for large language models to comprehend and reason about.
As with directly-follows graphs, it is important to note that there is not a single "correct" way to textually encode Petri nets. Advanced large language models are capable of understanding a wide range of textual encodings, provided the chosen method clearly communicates the structure and components of the Petri net. Different encodings might express the same Petri net, using various terminologies or formatting conventions to depict places, transitions, and arcs. As long as the encoding is clear and unambiguous, the language model will be able to understand and analyze the underlying Petri net effectively.
In addition, it is crucial to note that, unlike with directly-follows graphs, we cannot employ the same approach to addressing the context window limitation when dealing with Petri nets. The nature of Petri nets makes it necessary to abstract the entire model textually. The interconnected structure of places, transitions, and arcs in a Petri net forms a unified whole that cannot be easily segmented or reduced without potentially losing critical information or misrepresenting the process logic. Hence, the complete Petri net, regardless of its size, needs to be translated into text for the large language model to process effectively. This characteristic emphasizes the importance of developing efficient, concise and clear abstraction methods to avoid exceeding the context window of the large language model.

\begin{figure}[H]
\lstset{caption={Textual abstraction of a DECLARE process model computed on top of \autoref{table:traditionalEventLog} (obtained using the method \texttt{pm4py.llm.abstract\_declare}).}}
\lstset{label={lst:declareTextualAbstraction}}
\begin{lstlisting}
I have a DECLARE declarative process model containing the following constraints (here we provide a short explanation):
Existence: the activity is executed at least once.
Absence: the activity is not executed.
Exactly 1: the activity is executed exactly one time.
Initialization: the trace starts with one of the given activities.
Responded existence: given a couple of activities (A, B), if A occurs then B also occurs.
Co-Existence: given a couple of activities (A, B), if A occurs then B also occurs.
Response: given a couple of activities (A, B), if A occurs then B also occurs in the future of the trace.
Precedence: given a couple of activities (A, B), if B occurs then also A occurs in the past of the trace.
Succession: given a couple of activities (A, B), both the response and precedence constraints are satisfied.
Alternate response, alternate precedence, alternate succession: as the constraints mentioned above, but strenghtened by specifying that the events must alternate without repetitions.
Chain response, chain precedence, chain succession: as the constraints mentioned above, strenghtened by imposing the directly-follows relation.
Non Co-Existence: given a couple of activities (A, B), if A occurs then B should not occur.
Non Succession and non Chain succession: given a couple of activities (A, B), B should not follow A.

These are the constraints of the model:
Existence: Activity A
Exactly 1: Activity A
Initialization: Activity A
\end{lstlisting}
\end{figure}

When it comes to DECLARE process models (\autoref{subsec:declarativeProcModels}), a declarative process modeling notation, our approach to abstraction changes. These models are expressed as a set of constraints, each representing specific interdependencies and relationships between process activities. Given the non-trivial nature of these constraints, it becomes crucial to adequately explain them to the large language model.
A textual representation of a DECLARE model, as shown in \autoref{lst:declareTextualAbstraction}, translates these constraints into a readable format. However, unlike with procedural models, the form in which the constraints and explanations are presented is not rigid. Advanced large language models are capable of interpreting the constraints, regardless of how they're expressed, as long as the explanation remains clear and coherent.
However, one of the significant challenges when dealing with DECLARE models is the context window limitation. Real-life processes often involve hundreds of constraints, making their textual representations considerably long. To tackle this issue, one possible preprocessing strategy could be to rename the activities using alphabet letters or abbreviations. This approach considerably reduces the length of the textual encoding, ensuring it fits within the context window of the large language model, without losing the integrity of the model's information.

\begin{figure}[H]
\lstset{caption={Textual abstraction of temporal profile process model presented in \autoref{table:temporalProfile} (obtained using the method \texttt{pm4py.llm.abstract\_temporal\_profile}).}}
\lstset{label={lst:temporalProfileTextualAbstraction}}
\begin{lstlisting}
The temporal profile is a model describing the average and the standard deviation of the times between couples of activities eventually (not only directly) following each other in at least a process execution (so in a trace <A,B,C,D> the couples (A,B) (A,C) (A,D) (B,C) (B,D) (C,D) shall be considered). Given a positive value ZETA, a deviation occurs in a process execution when the time between two activities is lower than AVG - ZETA * STDEV or greater than AVG + ZETA * STDEV. For this process, the model is:

Activity A -> Activity B :  AVG: 1350.00 s  STD: 636.40 s
Activity A -> Activity C :  AVG: 1800.00 s  STD: 0.00 s
Activity B -> Activity C :  AVG: 900.00 s  STD: 0.00 s
Activity A -> Activity D :  AVG: 2700.00 s  STD: 0.00 s
Activity C -> Activity D :  AVG: 900.00 s  STD: 0.00 s
\end{lstlisting}
\end{figure}

In addition to the abstractions we've discussed, there are also temporal profile models to consider. A temporal profile model, another type of declarative process model (\autoref{subsec:declarativeProcModels}), is primarily concerned with the timing aspects of a process. The abstraction for such a model, as shown in \autoref{lst:temporalProfileTextualAbstraction}, encapsulates the temporal dependencies between various pairs of activities. Specifically, for each pair of activities that can potentially follow one another in the process, the abstraction provides the average and standard deviation of the time elapsed between them.
Just as with other models, advanced large language models are flexible in terms of the representation used for these constraints. As long as the pairs of activities and their corresponding time intervals are clearly stated, the large language model can interpret them, regardless of how they are presented in the abstraction. This versatility makes advanced large language models a potent tool for generating meaningful insights from abstracted temporal profile models.

\begin{figure}[H]
\lstset{caption={Textual abstraction of the OC-DFG presented in \autoref{fig:exampleOCDFG} (obtained using the method \texttt{pm4py.llm.abstract\_ocel\_ocdfg}).}}
\lstset{label={lst:ocdfgTextualAbstraction}}
\begin{lstlisting}
If I have an object-centric event log with the following directly follows graph (split between the different object types):

Object type: Order
  "Activity A" -> "Activity B" (frequency (number of events) = 3, frequency (number of objects) = 3, duration = 1200.00)
  "Activity C" -> "Activity A" (frequency (number of events) = 1, frequency (number of objects) = 1, duration = 1800.00)
  "Activity A" -> "Activity C" (frequency (number of events) = 1, frequency (number of objects) = 1, duration = 1800.00)
  "Activity C" -> "Activity D" (frequency (number of events) = 1, frequency (number of objects) = 1, duration = 900.00)

Object type: Invoices
  "Activity A" -> "Activity B" (frequency (number of events) = 3, frequency (number of objects) = 3, duration = 1200.00)
  "Activity C" -> "Activity A" (frequency (number of events) = 1, frequency (number of objects) = 1, duration = 1800.00)
  "Activity A" -> "Activity C" (frequency (number of events) = 1, frequency (number of objects) = 1, duration = 1800.00)
  "Activity C" -> "Activity D" (frequency (number of events) = 1, frequency (number of objects) = 1, duration = 900.00)

Object type: OrderItems
  "Activity A" -> "Activity B" (frequency (number of events) = 2, frequency (number of objects) = 3, duration = 900.00)
  "Activity A" -> "Activity D" (frequency (number of events) = 1, frequency (number of objects) = 1, duration = 2700.00)
  "Activity C" -> "Activity A" (frequency (number of events) = 1, frequency (number of objects) = 1, duration = 1800.00)
  "Activity A" -> "Activity C" (frequency (number of events) = 1, frequency (number of objects) = 1, duration = 1800.00)
\end{lstlisting}
\end{figure}

When it comes to object-centric event logs, we often utilize object-centric directly-follows graphs (\autoref{subsec:procedProcessModels}) as part of the abstraction process. An example of such an abstraction is depicted in \autoref{lst:ocdfgTextualAbstraction}. In this case, each type of object within the log has a corresponding set of arcs in its directly-follows graph. These arcs illustrate the flow and sequence of activities involving that particular object type.
However, similar to traditional directly-follows graphs, we must be mindful of the context window limitation. As a result, arcs with fewer occurrences may be omitted from the textual abstraction to maintain a manageable size. Regardless of this omission, the critical insight here is the pattern of activities and interactions concerning each object type.
Like the previously discussed abstractions, the way in which these object-specific arcs are represented in the abstraction does not significantly affect how advanced large language models interpret them. As long as the arcs are clearly stated, the large language model can efficiently process and generate useful insights from them. This flexibility underlines the applicability and utility of advanced large language models in understanding and analyzing complex object-centric artifacts.

\subsubsection{Textual Abstractions of Event Logs}
\label{subsubsec:textAbstrProMinEvLogs}

Process variants provide a crucial view of a traditional event log by giving us a glimpse into the different paths or sequences of activities that cases may follow in a process. They can be textually abstracted and presented to a Large Language Model (LLM), as illustrated in \autoref{lst:processVariantsTextualAbstraction}. Each variant is represented along with its frequency and average throughput time. This structure enables the LLM to understand both the variant's prevalence and the time typically required for its completion.
Given the frequency-based ordering of the variants, it is possible to handle limitations imposed by the context window. If the number of process variants is large and exceeds the context window size, less frequent variants can be omitted from the textual abstraction. Despite this omission, the abstraction would still retain the most common patterns, which likely capture a substantial portion of the overall process behavior.
The ability of advanced LLMs to understand different textual representations is particularly beneficial here. The representation of process variants does not have to adhere to a strict syntax. It could be a sequence of activities separated by symbols or a plain text description. 

\begin{figure}[H]
\lstset{caption={Textual abstraction of the process variants in \autoref{table:traditionalEventLog} (obtained using the method \texttt{pm4py.llm.abstract\_variants}).}}
\lstset{label={lst:processVariantsTextualAbstraction}}
\begin{lstlisting}
If I have a process with the following process variants:

 Activity A -> Activity C -> Activity D ( frequency = 1  performance = 2700.0 )
 Activity A -> Activity B -> Activity C ( frequency = 1  performance = 1800.0 )
 Activity A -> Activity B ( frequency = 1  performance = 1800.0 )
\end{lstlisting}
\end{figure}

When dealing with traditional event logs (\autoref{subsec:tradObjCentrEvLogs}), which are essentially organized collections of events grouped into cases, we can utilize textual abstractions to distill the key features of a specific case, as shown in \autoref{lst:caseTextualAbstraction}.
In this type of abstraction, we highlight the key attributes at the case level and provide a detailed list of all events associated with that case. Each event is principally characterized by an activity and a timestamp, but other event-level attributes can be included as necessary, providing a more detailed snapshot of each event.
The crucial aspect of this textual abstraction is that it presents all necessary information to evaluate the progression of a specific case. Consequently, it enables the large language model to answer questions that pertain to a specific case in question, such as identifying any potential issues or irregularities within that case. This makes case-level textual abstraction a powerful tool for detailed, case-specific analysis in process mining.

\begin{figure}[H]
\lstset{caption={Textual abstraction of \emph{Case 1} in \autoref{table:traditionalEventLog} (obtained using the method \texttt{pm4py.llm.abstract\_case}).}}
\lstset{label={lst:caseTextualAbstraction}}
\begin{lstlisting}
If I have a case with the following (case) attributes:

concept:name = Case1

the case contains the following events (the activity of the event is reported first):

Activity A ( timestamp = 2023-07-13 09:00:00+00:00 ;  @@case_index = 0 ;  @@index = 0 ; )
Activity B ( timestamp = 2023-07-13 09:15:00+00:00 ;  @@case_index = 0 ;  @@index = 1 ; )
Activity C ( timestamp = 2023-07-13 09:30:00+00:00 ;  @@case_index = 0 ;  @@index = 2 ; )
\end{lstlisting}
\end{figure}

A traditional event log can also be translated into a matrix of numerical features (\autoref{subsec:feaExt}), a format often used in machine learning. Each row in this matrix represents a case from the event log, and each column, or feature, provides some quantitative characteristic of that case. The textual abstraction of this type of matrix can be illustrated as in \autoref{lst:traditionalLogFeaturesTextualAbstraction}.
For every feature in the matrix, we provide its support, i.e., the number of cases where this feature is present or significant. Additionally, we supply quantiles of the feature values, giving a more detailed understanding of the distribution of these values. These aspects combined allow us to understand the relevance and descriptive power of each feature.
Given the potential breadth of this kind of matrix, it is essential to consider the context window limitation when producing a textual abstraction. As a solution, one could decide to omit features that have a low support. Such a choice would ensure that the most impactful features, those that occur more frequently, are represented in the abstraction, providing a succinct yet representative snapshot of the event log's numerical features. This strategy makes this type of textual abstraction extremely valuable for a broad understanding of the event log's characteristics from a numerical perspective.

\begin{figure}[H]
\lstset{caption={Textual abstraction of features extracted from the event log presented in \autoref{table:temporalProfile} (obtained using the method \texttt{pm4py.llm.abstract\_log\_features}).}}
\lstset{label={lst:traditionalLogFeaturesTextualAbstraction}}
\begin{lstlisting}
 Given the following features:

Value 'Activity A' for Event Attribute 'concept:name':    number of non-zero values: 3 ; quantiles of the non-zero: {0.0: 1.0, 0.25: 1.0, 0.5: 1.0, 0.75: 1.0, 1.0: 1.0}
Value 'Activity C' for Event Attribute 'concept:name':    number of non-zero values: 2 ; quantiles of the non-zero: {0.0: 1.0, 0.25: 1.0, 0.5: 1.0, 0.75: 1.0, 1.0: 1.0}
Value 'Activity B' for Event Attribute 'concept:name':    number of non-zero values: 2 ; quantiles of the non-zero: {0.0: 1.0, 0.25: 1.0, 0.5: 1.0, 0.75: 1.0, 1.0: 1.0}
Value 'Activity D' for Event Attribute 'concept:name':    number of non-zero values: 1 ; quantiles of the non-zero: {0.0: 1.0, 0.25: 1.0, 0.5: 1.0, 0.75: 1.0, 1.0: 1.0}
Succession 'Activity A' -> 'Activity B' for the Values of the Attribute 'concept:name':    number of non-zero values: 2 ; quantiles of the non-zero: {0.0: 1.0, 0.25: 1.0, 0.5: 1.0, 0.75: 1.0, 1.0: 1.0}
Succession 'Activity C' -> 'Activity D' for the Values of the Attribute 'concept:name':    number of non-zero values: 1 ; quantiles of the non-zero: {0.0: 1.0, 0.25: 1.0, 0.5: 1.0, 0.75: 1.0, 1.0: 1.0}
Succession 'Activity B' -> 'Activity C' for the Values of the Attribute 'concept:name':    number of non-zero values: 1 ; quantiles of the non-zero: {0.0: 1.0, 0.25: 1.0, 0.5: 1.0, 0.75: 1.0, 1.0: 1.0}
Succession 'Activity A' -> 'Activity C' for the Values of the Attribute 'concept:name':    number of non-zero values: 1 ; quantiles of the non-zero: {0.0: 1.0, 0.25: 1.0, 0.5: 1.0, 0.75: 1.0, 1.0: 1.0}
\end{lstlisting}
\end{figure}

Similar to the abstraction of traditional event logs, object-centric event logs can also be textually abstracted (\autoref{subsec:tradObjCentrEvLogs}), as exemplified in \autoref{lst:ocelComponentTextualAbstraction}. In an object-centric event log, each event is characterized not only by an activity and a timestamp, as in traditional event logs, but also by a set of related objects. These objects extend the dimensionality of the event log, providing more nuanced and contextually rich information about each event.
The textual abstraction of an object-centric event log essentially reflects this richer structure. Each event is paired with its associated objects, hence providing a thorough picture of the specific process instance that the event belongs to.

However, due to the typically large size of object-centric event logs, a direct abstraction of the entire log could easily exceed the context window limitation of the large language model. To address this issue, it is advisable to execute the textual abstraction against a specific process execution (as in \autoref{lst:ocelComponentTextualAbstraction}), which can be a connected component of related objects, rather than the entire log. This approach allows for localized question answering on a given process execution, ensuring that the resulting prompts are meaningful and contextually appropriate while staying within the context window limitation.
The ability to abstract and question specific process executions provides an extra level of flexibility when investigating process behaviours. It can help to unveil patterns or anomalies that are localized to specific instances of the process, contributing to a more granular understanding of the process dynamics.

\begin{figure}[H]
\lstset{caption={Textual abstraction of the connected component in \autoref{table:objectCentricEventLog} containing $\{ \textit{Order1}, \textit{Item1}, \textit{Item2}, \textit{Invoice1} \}$ (obtained using the method \texttt{pm4py.llm.abstract\_ocel}).}}
\lstset{label={lst:ocelComponentTextualAbstraction}}
\begin{lstlisting}
events:

Activity C ( related objects: Invoice2, Item3, Order2 )  timestamp: 2023-07-13 09:30:00
Activity A ( related objects: Invoice2, Item3, Order2 )  timestamp: 2023-07-13 10:00:00
Activity B ( related objects: Invoice2, Item4, Order2 )  timestamp: 2023-07-13 10:30:00

objects:

Invoice2 object type: Invoices ( lifecycle start: 2023-07-13 09:30:00 ; lifecycle end: 2023-07-13 10:30:00 ; lifecycle duration: 3600.0 )
Item3 object type: OrderItems ( lifecycle start: 2023-07-13 09:30:00 ; lifecycle end: 2023-07-13 10:00:00 ; lifecycle duration: 1800.0 )
Item4 object type: OrderItems ( lifecycle start: 2023-07-13 10:30:00 ; lifecycle end: 2023-07-13 10:30:00 ; lifecycle duration: 0.0 )
Order2 object type: Order ( lifecycle start: 2023-07-13 09:30:00 ; lifecycle end: 2023-07-13 10:30:00 ; lifecycle duration: 3600.0 )
\end{lstlisting}
\end{figure}

As with traditional event logs, object-centric event logs can also be transformed into a matrix of numerical features (\autoref{subsec:feaExt}), each representing a unique aspect of the process. These features often delve deeper, exploring the lifecycles of individual objects or interactions between them within the event log's graph structure. The textual abstraction of such features is presented in \autoref{lst:ocelFeaturesTextualAbstraction}, with each feature detailed by its support and quantiles of its values, mirroring the approach used for traditional logs.
However, due to the increased complexity and the larger number of features that object-centric event logs can offer, dealing with the context window limitation of Large Language Models can be more challenging. The primary strategy here is to focus on the most relevant features - those with the highest support. This prioritization helps to manage the amount of information fed into the LLM, ensuring that the most significant features are included, while those with less support can be omitted to fit within the context window. This approach allows for an effective trade-off between information detail and the size constraints imposed by the LLMs.

\begin{figure}[H]
\lstset{caption={Textual abstraction of the features for the 'Orders' object type of the object-centric event log presented in \autoref{table:objectCentricEventLog} (obtained using the method \texttt{pm4py.llm.abstract\_ocel\_features}).}}
\lstset{label={lst:ocelFeaturesTextualAbstraction}}
\begin{lstlisting}
Beforehand, a bit of notions. Given an object-centric event log, the object interaction graph connects objects that are related in at least an event. The object descendants graph connects objects related in at least an event, when the lifecycle of the second object starts after the lifecycle of the first. The object inheritance graph connects objects when there an event that ends the lifecycle of the first object and starts the lifecycle of the second one.
Given the following features:

Duration of the lifecycle of the object:    number of non-zero values: 4 ; quantiles of the non-zero: {0.0: 900.0, 0.25: 900.0, 0.5: 1800.0, 0.75: 2925.0, 1.0: 3600.0}
Number of unique activities in the lifecycle of the object:    number of non-zero values: 4 ; quantiles of the non-zero: {0.0: 2.0, 0.25: 2.0, 0.5: 2.5, 0.75: 3.0, 1.0: 3.0}
Number of events in the lifecycle of the object:    number of non-zero values: 4 ; quantiles of the non-zero: {0.0: 2.0, 0.25: 2.0, 0.5: 2.5, 0.75: 3.0, 1.0: 3.0}
Number of objects ending their lifecycle together with the current object:    number of non-zero values: 4 ; quantiles of the non-zero: {0.0: 2.0, 0.25: 2.0, 0.5: 2.0, 0.75: 2.25, 1.0: 3.0}
Number of objects starting their lifecycle together with the current object:    number of non-zero values: 4 ; quantiles of the non-zero: {0.0: 2.0, 0.25: 2.75, 0.5: 3.0, 0.75: 3.0, 1.0: 3.0}
Start timestamp of the lifecycle of the object:    number of non-zero values: 4 ; quantiles of the non-zero: {0.0: 1689238800.0, 0.25: 1689240150.0, 0.5: 1689243300.0, 0.75: 1689246900.0, 1.0: 1689249600.0}
Completion timestamp of the lifecycle of the object:    number of non-zero values: 4 ; quantiles of the non-zero: {0.0: 1689239700.0, 0.25: 1689243075.0, 0.5: 1689246450.0, 0.75: 1689249150.0, 1.0: 1689250500.0}
Number of occurrences of the activity Activity A:    number of non-zero values: 4 ; quantiles of the non-zero: {0.0: 1.0, 0.25: 1.0, 0.5: 1.0, 0.75: 1.0, 1.0: 1.0}
Number of objects related in the object interaction graph:    number of non-zero values: 4 ; quantiles of the non-zero: {0.0: 3.0, 0.25: 3.0, 0.5: 3.0, 0.75: 3.0, 1.0: 3.0}
Number of object of type OrderItems related to the current object in the object interaction graph:    number of non-zero values: 4 ; quantiles of the non-zero: {0.0: 2.0, 0.25: 2.0, 0.5: 2.0, 0.75: 2.0, 1.0: 2.0}
Number of object of type Invoices related to the current object in the object interaction graph:    number of non-zero values: 4 ; quantiles of the non-zero: {0.0: 1.0, 0.25: 1.0, 0.5: 1.0, 0.75: 1.0, 1.0: 1.0}
Number of occurrences of the activity Activity B:    number of non-zero values: 3 ; quantiles of the non-zero: {0.0: 1.0, 0.25: 1.0, 0.5: 1.0, 0.75: 1.0, 1.0: 1.0}
Frequency of the path "Activity A" -> "Activity B" in the lifecycle of the object:    number of non-zero values: 3 ; quantiles of the non-zero: {0.0: 1.0, 0.25: 1.0, 0.5: 1.0, 0.75: 1.0, 1.0: 1.0}
Number of occurrences of the activity Activity C:    number of non-zero values: 2 ; quantiles of the non-zero: {0.0: 1.0, 0.25: 1.0, 0.5: 1.0, 0.75: 1.0, 1.0: 1.0}
Number of occurrences of the activity Activity D:    number of non-zero values: 1 ; quantiles of the non-zero: {0.0: 1.0, 0.25: 1.0, 0.5: 1.0, 0.75: 1.0, 1.0: 1.0}
Number of objects which precede the current object in the object descendants graph:    number of non-zero values: 1 ; quantiles of the non-zero: {0.0: 1.0, 0.25: 1.0, 0.5: 1.0, 0.75: 1.0, 1.0: 1.0}
Number of objects which follow the current object in the object inheritance graph:    number of non-zero values: 1 ; quantiles of the non-zero: {0.0: 1.0, 0.25: 1.0, 0.5: 1.0, 0.75: 1.0, 1.0: 1.0}
Number of objects which follow the current object in the object descendants graph:    number of non-zero values: 1 ; quantiles of the non-zero: {0.0: 1.0, 0.25: 1.0, 0.5: 1.0, 0.75: 1.0, 1.0: 1.0}
\end{lstlisting}
\end{figure}

\subsection{Prompting Strategies}
\label{subsec:promptingStrategies}

This section introduces three prompting strategies for utilizing large language models (LLMs) in the field of process mining: "direct answering," "multi-prompt answering," and "generation of database queries." These strategies vary in complexity and are chosen based on the problem at hand. They make use of textual abstractions and questions, and leverage the capabilities of LLMs for hypothesis generation, result interpretation, and even SQL query formulation, providing an interactive and adaptive approach to process mining.

\subsubsection{Direct Answering}
\label{subsubsec:strategyDirectAnswering}

The "direct answering" strategy is the most straightforward prompting approach used with large language models in the context of process mining. This strategy involves formulating a prompt which combines a textual abstraction of a process mining artifact with a direct question related to it. Example questions might be "What are the bottlenecks of the process?" (starting from the process variants abstraction) or "What are the main anomalies?" (starting from the directly-follows graph abstraction).

It is important to note that, in this strategy, the output structure of the response is not predetermined. Rather, the large language model will generate a response in the form of free text. Consequently, this means that the user, while freed from the task of analyzing the event log themselves, must have sufficient background knowledge to interpret the answer correctly.

Another key factor for success with the direct answering strategy is the formulation of the question itself. It needs to be concise, clear, and well-structured to enable the large language model to provide a meaningful and relevant response. This strategy, therefore, represents a delicate balance between user expertise, question formulation, and the interpretability of the model's responses.

\subsubsection{Multi-Prompt Answering}
\label{subsubsec:strategyMultiPromptAnswering}

In certain scenarios, more complex process mining tasks require a series of distinct prompts, each involving different abstractions and questions. This approach, known as multi-prompt answering, leverages the iterative nature of inquiry, where the answer to one question informs the formulation of the next. In this way, the user guides the large language model (LLM) through a sequence of steps to ultimately resolve the overarching process mining problem.

A classic example where the multi-prompt answering strategy is useful is in assessing fairness in process mining. Initially, a user might use a feature table abstraction to identify attributes that could potentially lead to discrimination. Once the LLM provides this information, the user formulates a subsequent prompt to investigate if the process execution differs for cases belonging to the protected and non-protected groups. The second prompt is contingent on the results from the first one, as it is necessary to distinguish between cases before delving deeper into potential discrepancies in process execution.

In another example, the user might want to analyze process conformance. An initial prompt could be used to compare the process model to a log abstraction to identify non-conforming activities. Subsequent prompts could then be designed to examine why these non-conformities are occurring, perhaps by examining specific cases or focusing on particular process variants.

Multi-prompt answering is a powerful strategy, as it allows users to conduct a step-by-step, deep-dive analysis. However, it requires careful formulation of prompts and the ability to interpret and act on the LLM responses to guide the inquiry effectively.
Therefore, it demands more user engagement than the direct answering strategy.

\subsubsection{Generation of Database Queries}
\label{subsubsec:strategyGenerationDbQueries}

The third strategy is "generation of database queries", which is particularly useful for complex process mining problems that are too intricate to be directly abstracted or solved by the large language model (LLM). Questions like "What is the average throughput time when activity B occurs more than one day after the start of the case?" or "Measure the correlation between throughput time and rework in the case" fall into this category.

In these situations, the LLM can be leveraged to generate SQL queries based on the user's initial natural language inquiry. These generated queries can then be executed against the event log, enabling complex data analysis without the user having to manually construct the intricate SQL statements.

Yet, this strategy's potential goes beyond translating natural language questions into SQL queries. It can also be used to generate hypotheses about the event log, derived from certain abstractions like process variants or log features. For instance, the user can present the LLM with such an abstraction and ask it to formulate an hypothesis about the event log expressed as a SQL query.

Once the SQL query (representing the hypothesis) is executed and the results are obtained, the LLM can then be employed again to evaluate if the original hypothesis holds true. Should the initial hypothesis be invalidated, the LLM is capable of formulating alternative hypotheses, which can then be transformed into SQL queries and tested against the data.

This iterative method, which encompasses generating hypotheses, testing them, and re-assessing based on the outcomes, allows for a more in-depth exploration and understanding of the underlying processes. This approach significantly expands the scope of possible investigations that can be conducted using the LLM.

Nevertheless, while this method can streamline complex query generation and hypothesis testing, it does call for a specific set of skills on the part of the user.
While the LLM can assist in generating complex queries and hypotheses, the responsibility of result interpretation and subsequent steps of analysis falls on the user. Thus, a certain degree of analytical acumen and domain knowledge is necessary for this approach to be fully effective.

\section{Tool Support}
\label{sec:toolSupport}

In this section, we delve into the tool support available for conducting process mining tasks with large language models.
\emph{pm4py} \cite{pm4py}, short for Process Mining for Python, is a versatile and comprehensive library that supports the entire spectrum of process mining applications, ranging from data loading and pre-processing, to process discovery, conformance checking, and enhancement. This suite of functionality enables practitioners and researchers to extract valuable insights from event logs and use them for optimizing and monitoring real-world processes.
pm4py is developed from the Fraunhofer FIT process mining group and is fully documented at the address \url{https://pm4py.fit.fraunhofer.de}. 

\begin{table}[ht]
\centering
\begin{tabular}{|p{0.45\linewidth}|p{0.45\linewidth}|}
\hline
\textbf{Method} & \textbf{Abstraction Provided} \\
\hline
\texttt{pm4py.llm.abstract\_dfg()} & Provides the DFG abstraction of a traditional event log \\
\texttt{pm4py.llm.abstract\_variants()} & Provides the variants abstraction of a traditional event log \\
\texttt{pm4py.llm.abstract\_log\_attributes()} & Provides the abstraction of the attributes/columns of the event log \\
\texttt{pm4py.llm.abstract\_log\_features()} & Provides the abstraction of the machine learning features obtained from an event log \\
\texttt{pm4py.llm.abstract\_case()} & Provides the abstraction of a case (collection of events) \\
\texttt{pm4py.llm.abstract\_ocel()} & Provides the abstraction of an object-centric event log (list of events and objects) \\
\texttt{pm4py.llm.abstract\_ocel\_ocdfg()} & Provides the abstraction of an object-centric event log (OC-DFG) \\
\texttt{pm4py.llm.abstract\_ocel\_features()} & Provides the abstraction of an object-centric event log (features for ML) \\
\texttt{pm4py.llm.abstract\_event\_stream()} & Provides an abstraction of the (last) events of the stream related to a traditional event log \\
\texttt{pm4py.llm.abstract\_petri\_net()} & Provides the abstraction of a Petri net \\
\texttt{pm4py.llm.abstract\_log\_skeleton()} & Provides the abstraction of a log skeleton model \\
\hline
\end{tabular}
\caption{Methods for various types of abstractions available in \emph{pm4py}.}
\label{table:abstractionMethods}
\end{table}

A significant development in version 2.7.5 of pm4py is the integration with Large Language Models (LLM), bolstering the tool's capacity for analysis and understanding of process data. This integration enables pm4py to offer a range of abstractions for process data and models, as detailed in \autoref{table:abstractionMethods}. More features will be introduced in pm4py 2.7.6. Each method in this module provides a different view or representation of the process data, facilitating diverse avenues for exploration and analysis.

\begin{table}[ht]
\centering
\begin{tabular}{|p{0.45\linewidth}|p{0.45\linewidth}|}
\hline
\textbf{Method} & \textbf{Functionality} \\
\hline
\texttt{pm4py.llm.openai\_query()} & Executes a prompt against OpenAI, returning the response as string \\
\hline
\end{tabular}
\caption{Method for interacting directly with the LLM APIs available in \emph{pm4py}.}
\label{table:apiMethod}
\end{table}

In addition to these abstractions, pm4py 2.7.5 allows for direct interaction with LLM through specific methods such as \texttt{pm4py.llm.openai\_query} (\autoref{table:apiMethod}). This function enables users to directly execute prompts against OpenAI's language models, obtaining textual responses that can be further analyzed or used in subsequent processes.

\begin{figure}[H]
\lstset{language=Python}
\lstset{frame=lines}
\lstset{caption={Example usage of the pm4py's LLM integration on traditional process mining objects}}
\lstset{label={lst:exampleUsageTraditionalObjects}}
\lstset{basicstyle=\scriptsize}
\begin{lstlisting}
import pm4py
log = pm4py.read_xes("tests/input_data/roadtraffic100traces.xes")
q1 = """\n What are the root causes of the performance issues in the process?
Please provide only process and data specific considerations,
no general considerations."""
print(pm4py.llm.abstract_variants(log) + q1)

net, im, fm = pm4py.discover_petri_net_inductive(log)
q2 = """\n Can you provide suggestions to improve the process model
based on your domain knowledge?"""
print(pm4py.llm.abstract_petri_net(net, im, fm) + q2)
\end{lstlisting}
\end{figure}

A sample usage of pm4py's LLM integration is shown in \autoref{lst:exampleUsageTraditionalObjects}, where the \texttt{abstract\_variants} method is used to retrieve the variants of an event log. Subsequently, the Petri net discovered from the event log is abstracted using \texttt{abstract\_petri\_net}, and both of these abstractions are used to construct prompts for querying an LLM.

\begin{figure}[H]
\lstset{language=Python}
\lstset{frame=lines}
\lstset{caption={Example usage of the pm4py's LLM integration on object-centric process mining objects}}
\lstset{label={lst:exampleUsageOCELObjects}}
\lstset{basicstyle=\scriptsize}
\begin{lstlisting}
import pm4py
ocel = pm4py.read_ocel("tests/input_data/ocel/example_log.jsonocel")

q1 = "\n\nWhat are the strangest relationships in this graph?"
print(pm4py.llm.abstract_ocel_ocdfg(ocel) + q1)

q2 = "\n\nWhat are the strangest features' values?"
print(pm4py.llm.abstract_ocel_features(ocel, "order") + q2)
\end{lstlisting}
\end{figure}

A distinct advantage of the pm4py tool is its support for object-centric process mining, a relatively recent advancement in the process mining field that provides a more detailed view of complex processes. Object-centric process mining expands the traditional case-centric perspective, allowing for the consideration of multiple interacting objects in a process, thus enabling a more nuanced understanding of process dynamics.

The LLM integration in pm4py embraces this object-centric perspective, as demonstrated in the example given in \autoref{lst:exampleUsageOCELObjects}. The script loads an object-centric event log (OCEL) and then uses two specific abstraction methods tailored for this type of data.
The method \texttt{abstract\_ocel\_ocdfg} is used to generate an abstraction of the Object-Centric Directly Follows Graph (OC-DFG), a representation that captures the relationships between the different activities and objects in the log. This abstraction is then used to create a query for the LLM, asking for unusual relationships in the process graph.
The method \texttt{abstract\_ocel\_features} is used to generate an abstraction of the features of a specific type of object (in this example, "order") in the OCEL. This abstraction is then used to create another query for the LLM, asking for odd feature values.

These examples showcase the potential of using LLM in conjunction with object-centric process mining. The intricate process insights offered by object-centric logs, coupled with the rich interpretative capabilities of large language models, provide practitioners and researchers with powerful tools to uncover hidden process knowledge and drive informed process decisions.

\section{Assessment}
\label{sec:assessment}

In this section, we examine the ability of large language models, specifically Bard and GPT-4, to understand, generate, and answer queries related to process mining tasks. The considered models and their unique characteristics are discussed in \autoref{subsec:assessmentConsideredLLMs}. We also outline the specific event logs leveraged in this assessment in \autoref{subsec:assessmentConsideredEventLogs}.

We delve into the models' capabilities in the realm of direct answering to process mining queries in \autoref{sec:assessmentDirectAnswering}, where we assess their ability to respond directly to various types of prompts. This exploration is then extended in \autoref{sec:assessmentMultiPromptAnswering}, where we investigate the models' capacity for providing answers across multiple prompts, encapsulating more complex interactions.

Finally, we evaluate the models' aptitude in generating database queries from user prompts in \autoref{sec:assessmentGenerationDatabaseQueries}. This involves assessing the ability of these models to not just understand a prompt, but to translate that understanding into SQL queries, a crucial aspect in the context of data analysis.

By comprehensively assessing these aspects, we aim to gauge the practicality and potential of these large language models in process mining tasks, seeking to illuminate areas of strength as well as identify opportunities for further research and improvement.

\subsection{Considered Large Language Models}
\label{subsec:assessmentConsideredLLMs}

\begin{table}[ht]
\caption{Fundamental statistics of the real-life event logs used in the assessment.}
\resizebox{\textwidth}{!}{
\begin{tabular}{lrrrrr}
\toprule
{\bf Event Log} & {\bf Number of Events} & {\bf Number of Cases} & {\bf Number of Variants} & {\bf Number of Activities} \\
\midrule
Road Traffic & 561470 & 150370 & 231 & 11 \\
BPI Challenge 2020 & 56437 & 10500 & 99 & 17 \\
Sepsis & 15214 & 1050 & 846 & 16 \\
Conformance Checking Challenge 2019 & 1394 & 20 & 20 & 29 \\
\bottomrule
\end{tabular}
}
\label{tbl:eventLogsRealLife}
\end{table}

This section provides an examination of the two advanced LLMs currently leveraged in the field of process mining: GPT-4, developed by OpenAI, and Google Bard, brought forward by Google AI. Both models are the product of extensive training on vast datasets comprised of a diverse range of text and code, and exhibit remarkable capabilities in a variety of tasks.

\emph{GPT-4} is a large language model (LLM) developed by OpenAI. It is a powerful tool that can be used for a variety of tasks, including text generation, translation, and question answering. GPT-4 is trained on a massive dataset of text and code, which allows it to generate text that is both coherent and grammatically correct.

\emph{Google Bard} is a large language model developed by Google AI. It is similar to GPT-4 in terms of its capabilities, but it is trained on a different dataset of text and code. This difference in training data means that Google Bard has different strengths and weaknesses than GPT-4.

\subsection{Considered Event Logs}
\label{subsec:assessmentConsideredEventLogs}

In this study, we focus on several real-world process event logs to explore and evaluate the capabilities of large language models in the domain of process mining. These event logs encapsulate diverse fields, such as road traffic fine management, hospital case management, and business travel expenses, offering a breadth of scenarios to test the adaptability and accuracy of the language model. The event logs employed in this research are described in detail below.

\begin{table}[ht]
\caption{Fundamental statistics of the event logs used for fairness evaluation.}
\resizebox{\textwidth}{!}{
\begin{tabular}{lrrrrr}
\toprule
{\bf Event Log} & {\bf Number of Events} & {\bf Number of Cases} & {\bf Number of Variants} & {\bf Number of Activities} \\
\midrule
hiring\_log\_high-xes.gz & 63869 & 10000 & 386 & 12 \\
hiring\_log\_low-xes.gz & 72094 & 10000 & 296 & 12 \\
hiring\_log\_medium-xes.gz & 69054 & 10000 & 382 & 12 \\
\hline
hospital\_log\_high-xes.gz & 69528 & 10000 & 80 & 10 \\
hospital\_log\_low-xes.gz & 70037 & 10000 & 106 & 10 \\
hospital\_log\_medium-xes.gz & 70124 & 10000 & 77 & 10 \\
\hline
lending\_log\_high-xes.gz & 58822 & 10000 & 41 & 12 \\
lending\_log\_low-xes.gz & 60746 & 10000 & 31 & 12 \\
lending\_log\_medium-xes.gz & 58668 & 10000 & 33 & 12 \\
\hline
renting\_log\_high-xes.gz & 89972 & 10000 & 496 & 16 \\
renting\_log\_low-xes.gz & 96440 & 10000 & 508 & 16 \\
renting\_log\_medium-xes.gz & 105555 & 10000 & 610 & 16 \\
\bottomrule
\end{tabular}
}
\label{tbl:eventLogsFairness}
\end{table}

\begin{table}[ht]
\caption{Sensitive attributes in the supplied event log dataset.}
\resizebox{\textwidth}{!}{
\begin{tabular}{|l|c|c|c|c|c|c|c|c|}
\hline
\textbf{Domain} & \textbf{Age} & \textbf{Citizenship} & \textbf{German Proficiency} & \textbf{Gender} & \textbf{Religion} & \textbf{Years of Education} & \textbf{Underlying Condition} & \textbf{Private Insurance} \\
\hline
Hiring & X & X & X & X & X & X & ~ & ~ \\
\hline
Hospital & X & X & X & X & ~ & ~ & X & X \\
\hline
Lending & X & X & X & X & ~ & X & ~ & ~ \\
\hline
Renting & X & X & X & X & X & X & ~ & ~ \\
\hline
\end{tabular}
}
\label{tbl:sensitiveAttributes}
\end{table}

\begin{itemize}
\item \emph{Road Traffic Fine Management Process}\footnote{\url{https://doi.org/10.4121/uuid:270fd440-1057-4fb9-89a9-b699b47990f5}}: real-life event log of an information system managing road traffic fines in an Italian municipality.
\item \emph{BPI Challenge 2020 (Domestic Declarations)}\footnote{\url{https://doi.org/10.4121/uuid:3f422315-ed9d-4882-891f-e180b5b4feb5}}: the travel expense process consists of an employee's submission, approval by the travel administration, then by the budget owner and supervisor, and possibly a director. The process concludes with a trip or a payment request. There are two types of trips: domestic and international. Domestic trips do not need prior approval, allowing employees to ask for cost reimbursement afterward. 
\item \emph{Sepsis Cases Event Log}\footnote{\url{https://data.4tu.nl/articles/_/12707639/1}}: the dataset is an anonymized real-life event log of sepsis cases from a hospital, where sepsis is a life-threatening condition often resulting from an infection. Each case represents a patient's journey through the hospital, with events captured by the hospital's Enterprise Resource Planning (ERP) system. The events cover a range of different activities, and a multitude of data attributes are recorded such as the group responsible for the activity, test results, and checklist information. While event timestamps have been randomized, the time interval between events within a case remains unchanged.
\item \emph{Conformance Checking Challenge 2019}\footnote{\url{https://doi.org/10.4121/uuid:c923af09-ce93-44c3-ace0-c5508cf103ad}}: This past version of the conformance checking challenge was centered around a process involved in medical training. Specifically, it revolved around how medical students were taught to place a central venous catheter (CVC) using ultrasound guidance. The CVC procedure involves inserting a tube into a large vein, a process critical for administering fluids or medicines to patients, among other applications.
\end{itemize}

For assessment of the multi-prompt answering strategy, we also use the event logs provided in \cite{pohl2023collection} and described in \cite{DBLP:journals/corr/abs-2306-11453}, including:
\begin{itemize}
\item \emph{Hiring event logs}: these logs chronicle a complex recruitment process with various stages of evaluation and decision-making. The applicant's journey can be as simple as a straightforward rejection or as elaborate as multi-step screening culminating in a job offer, contingent on a host of influencing factors.
\item \emph{Hospital event logs}: this dataset portrays a patient's journey through a hospital system, starting from the initial check-in at the Emergency Room or Family Department. It progresses through phases of medical examination, diagnosis, and treatment. It's worth noting that unsuccessful treatments often necessitate repeat examinations and interventions, illustrating the cyclical nature of healthcare delivery.
\item \emph{Lending event logs}: these logs detail a loan application process starting from the scheduling of an initial appointment. It traverses several steps including validation of provided information and underwriting, leading to either loan approval or denial. Further requirements, such as the involvement of a co-signer or an assessment of collateral, may be needed. Some cases may experience direct appointment refusal, highlighting the process's variability based on applicants' individual financial circumstances.
\item \emph{Renting event logs}: these logs present a comprehensive rental process that begins with a property viewing request and can end with a contract termination. Between these endpoints, stages such as preliminary vetting, property viewing, decision-making, and possibly, intensive screening occur. If successful, a rental agreement is initiated, with possible outcomes ranging from eviction due to late payments to voluntary contract termination by tenants. It's important to note that not all applications advance to the property viewing phase.
\end{itemize}
The degree of discrimination exhibited in the logs varies across different domains, offering researchers an opportunity to explore tangible real-world situations. Table \ref{tbl:eventLogsFairness} furnishes fundamental statistics for the log collection, and Table \ref{tbl:sensitiveAttributes} elaborates on the sensitive attributes. It's important to note that the selection of attributes pertaining to fairness can be a subject of debate; hence, we welcome discussions to enhance our grasp of fairness within the realm of process mining. Detailed descriptions of each log's attributes and process have been provided to facilitate the identification of potential instances of discrimination.
The event logs provided also include the ground truth concerning the classification of cases - whether they belong to the protected group or not. This inclusion enables a comparison between actual case classifications, which are based on the sensitive attributes, and the actual ground truth. Therefore, the capacity to evaluate the accuracy of case categorization into protected and non-protected groups is significantly enhanced.

In our exploration of object-centric process mining using large language models, we utilized several publicly accessible logs available at \url{https://www.ocel-standard.org}. Specifically, two simulated object-centric event logs were chosen for this analysis:
\begin{itemize}
\item the 'Order Management' event log, which can be accessed directly at \url{https://www.ocel-standard.org/1.0/running-example.jsonocel.zip}. This log contains the object types 'orders', 'items', and 'packages', all of which were retained for our assessment.
\item the 'Recruiting' event log, available at \url{https://www.ocel-standard.org/1.0/recruiting.jsonocel.zip}. In this log, we focused on the 'applications' and 'offers' object types.
\end{itemize}

By leveraging these specific logs, we aimed to evaluate how effectively large language models like Bard and GPT-4 could handle complex, object-centric process mining tasks.

\subsection{Direct Answering}
\label{sec:assessmentDirectAnswering}

In this section, we conduct an evaluation of the ability of advanced large language models, specifically Bard and GPT-4, to respond effectively to process mining queries, given the required abstraction. Our assessments of Bard and GPT-4 with regards to traditional process mining and event logs are documented in \autoref{tab:traditionalEventLogAssessment}.

In the table, each query is assigned a unique code, and the appropriate acceptance criteria for the responses generated by the large language models are stipulated. Furthermore, each query is examined across four distinct event logs, namely \emph{Road Traffic}, \emph{BPI Challenge 2020}, \emph{Sepsis}, and \emph{CCC19}.

Observably, both Bard and GPT-4 demonstrate commendable performance when dealing with descriptive queries such as {\bf TQ1}, {\bf TQ2}, {\bf TQ3}, and {\bf TQ5}. Furthermore, these large language models exhibit a fair understanding of procedural and declarative models, demonstrating capabilities in tasks like play-out and conformance checking, albeit with certain constraints.

In particular, GPT-4 displayed a satisfactory level of competency across the diverse range of queries, albeit with some restrictions pertaining to medical processes. Therefore, it is suggested that further refinement and training might be beneficial in enhancing the capability of these models in the realm of process mining.

\begin{landscape}
\begin{table}[ht]
\centering
\caption{Assessment of large language models' capabilities on traditional event logs/process models.}
\resizebox{18cm}{!}{
\begin{tabular}{|p{7cm}|p{7cm}|p{4cm}|p{2cm}|p{2cm}|p{2cm}|p{2cm}|p{2cm}|p{2cm}|p{2cm}|p{2cm}|}
\hline
\textbf{Question} & \textbf{Acceptance Criteria for the Answer(s)} & \textbf{Textual Abstraction Used} & \multicolumn{2}{c|}{\textbf{Road Traffic}} & \multicolumn{2}{c|}{\textbf{BPI Challenge 2020}} & \multicolumn{2}{c|}{\textbf{Sepsis}} & \multicolumn{2}{c|}{\textbf{CCC19}} \\
\hline
 &  &  & \textbf{GPT-4} & \textbf{Bard} & \textbf{GPT-4} & \textbf{Bard} & \textbf{GPT-4} & \textbf{Bard} & \textbf{GPT-4} & \textbf{Bard} \\
\hline
{\bf TQ1} Could you provide a description of the process underlying this data? & The description reflects the process and its steps. & DFG & \textcolor{cadmiumgreen}{$\surd$} & \textcolor{cadmiumgreen}{$\surd$} &  \textcolor{cadmiumgreen}{$\surd$} & \textcolor{cadmiumgreen}{$\surd$} & \textcolor{cadmiumgreen}{$\surd$} & \textcolor{cadmiumgreen}{$\surd$} & \textcolor{cadmiumgreen}{$\surd$} & \textcolor{cadmiumgreen}{$\surd$}  \\
\hline
{\bf TQ2} Could you provide a detailed description of the most frequent activity of the process? & The most frequent activity is identified and described thoroughly. & DFG & \textcolor{cadmiumgreen}{$\surd$} & \textcolor{cadmiumgreen}{$\surd$} & \textcolor{cadmiumgreen}{$\surd$} & \textcolor{cadmiumgreen}{$\surd$} & \textcolor{cadmiumgreen}{$\surd$} & \textcolor{cadmiumgreen}{$\surd$} & \textcolor{cadmiumgreen}{$\surd$} & \textcolor{cadmiumgreen}{$\surd$} \\
\hline
{\bf TQ3} Could you provide a detailed description of the most frequent path of the process? & The most frequent path is identified and described thoroughly. & DFG & \textcolor{cadmiumgreen}{$\surd$} & \textcolor{cadmiumgreen}{$\surd$} & \textcolor{cadmiumgreen}{$\surd$} & \textcolor{cadmiumgreen}{$\surd$} & \textcolor{cadmiumgreen}{$\surd$} & \textcolor{cadmiumgreen}{$\surd$} & \textcolor{cadmiumgreen}{$\surd$} & \textcolor{orange}{$\surd$} \\
\hline
{\bf TQ4} Could you identify the main anomalies in this data according to your domain knowledge of the process? & Some anomalies are correctly identified from the data. & DFG & \textcolor{orange}{$\surd$} & \textcolor{red}{X} & \textcolor{cadmiumgreen}{$\surd$} & \textcolor{orange}{$\surd$} & \textcolor{cadmiumgreen}{$\surd$} & \textcolor{red}{X} & \textcolor{orange}{$\surd$} & \textcolor{red}{X} \\
\hline
{\bf TQ5} Could you provide a detailed description of the most frequent process variant of the process? & The most frequent process variant is identified and described thoroughly. & Process Variants & \textcolor{cadmiumgreen}{$\surd$} & \textcolor{cadmiumgreen}{$\surd$} & \textcolor{cadmiumgreen}{$\surd$} & \textcolor{cadmiumgreen}{$\surd$} & \textcolor{cadmiumgreen}{$\surd$} & \textcolor{cadmiumgreen}{$\surd$} & \textcolor{cadmiumgreen}{$\surd$} & \textcolor{orange}{$\surd$} \\
\hline
{\bf TQ6} Could you identify the root causes of the performance issues in the process? & Some root causes for performance issues are correctly identified from the data. & Process Variants & \textcolor{orange}{$\surd$} & \textcolor{red}{X} & \textcolor{cadmiumgreen}{$\surd$} & \textcolor{red}{X} & \textcolor{red}{X} & \textcolor{red}{X} & \textcolor{orange}{$\surd$} & \textcolor{red}{X} \\
\hline
{\bf TQ7} Can you provide me a trace allowed by the process model? & The provided trace is allowed from the procedural model. & Petri net & \textcolor{cadmiumgreen}{$\surd$} & \textcolor{red}{X} & \textcolor{cadmiumgreen}{$\surd$} & \textcolor{red}{X} & \textcolor{red}{X} & \textcolor{red}{X} & \textcolor{red}{X} & \textcolor{red}{X} \\
\hline
{\bf TQ8} Can you explain why this trace is not allowed by the process model? & A valid explanation about the unconformity is provided. & Petri net & \textcolor{cadmiumgreen}{$\surd$} & \textcolor{cadmiumgreen}{$\surd$} & \textcolor{cadmiumgreen}{$\surd$} & \textcolor{red}{X} & \textcolor{orange}{$\surd$} & \textcolor{red}{X} & \textcolor{red}{X} & \textcolor{red}{X} \\
\hline
{\bf TQ9} Can you provide me a trace allowed by the process model? & The provided trace is allowed from the declarative model. & DECLARE & \textcolor{cadmiumgreen}{$\surd$} & \textcolor{cadmiumgreen}{$\surd$} & \textcolor{cadmiumgreen}{$\surd$} & \textcolor{red}{X} & \textcolor{red}{X} & \textcolor{red}{X} & ~ & ~ \\
\hline
{\bf TQ10} Can you explain why this trace is not allowed by the process model? & A valid explanation about the unconformity is provided. & DECLARE & \textcolor{cadmiumgreen}{$\surd$} & \textcolor{cadmiumgreen}{$\surd$} & \textcolor{cadmiumgreen}{$\surd$} & \textcolor{orange}{$\surd$} & \textcolor{cadmiumgreen}{$\surd$} & \textcolor{red}{X} & ~ & ~ \\
\hline
{\bf TQ11} Can you explain why this trace is not allowed by the process model? & A valid explanation about the unconformity is provided. & Temporal Profile & \textcolor{cadmiumgreen}{$\surd$} & \textcolor{cadmiumgreen}{$\surd$} & \textcolor{red}{X} & \textcolor{red}{X} & \textcolor{red}{X} & \textcolor{red}{X} & ~ & ~ \\
\hline
{\bf TQ12} Could you identify interesting aspects for the analysis of the process? & Some aspects to investigate are correctly identified from the data. & Log Features & \textcolor{cadmiumgreen}{$\surd$} & \textcolor{red}{X} & \textcolor{orange}{$\surd$} & \textcolor{red}{X} & \textcolor{cadmiumgreen}{$\surd$} & \textcolor{red}{X} & \textcolor{red}{X} & \textcolor{red}{X} \\
\hline
\end{tabular}
}
\label{tab:traditionalEventLogAssessment}
\end{table}
\begin{table}[ht]
\centering
\caption{Assessment of large language models' capabilities on object-centric event logs.}
\resizebox{18cm}{!}{
\begin{tabular}{|p{7cm}|p{7cm}|p{4cm}|p{2cm}|p{2cm}|p{2cm}|p{2cm}|}
\hline
\textbf{Question} & \textbf{Acceptance Criteria for the Answer(s)} & \textbf{Textual Abstraction Used} & \multicolumn{2}{c|}{\textbf{Order Management}} & \multicolumn{2}{c|}{\textbf{Recruiting}} \\
\hline
 &  &  & \textbf{GPT-4} & \textbf{Bard} & \textbf{GPT-4} & \textbf{Bard} \\
\hline
{\bf OQ1} Could you provide a description of the process underlying this data? & The description reflects the process and its steps. & OC-DFG & \textcolor{cadmiumgreen}{$\surd$} & \textcolor{cadmiumgreen}{$\surd$} & \textcolor{cadmiumgreen}{$\surd$} & \textcolor{orange}{$\surd$} \\
\hline
{\bf OQ2} Could you identify the main anomalies in this data according to your domain knowledge of the process? & Some anomalies are correctly identified from the data. & OC-DFG & \textcolor{cadmiumgreen}{$\surd$} & \textcolor{orange}{$\surd$} & \textcolor{cadmiumgreen}{$\surd$} & \textcolor{orange}{$\surd$} \\
\hline
{\bf OQ3} Could you identify the root causes of the performance issues in the process? & Some root causes for performance issues are correctly identified from the data. & OC-DFG & \textcolor{orange}{$\surd$} & \textcolor{cadmiumgreen}{$\surd$} & \textcolor{cadmiumgreen}{$\surd$} & \textcolor{red}{X} \\
\hline
{\bf OQ4} Could you identify the main anomalies in this data according to your domain knowledge of the process? & Some anomalies are correctly identified from the data. & Process Execution & \textcolor{cadmiumgreen}{$\surd$} & \textcolor{cadmiumgreen}{$\surd$} & \textcolor{cadmiumgreen}{$\surd$} & \textcolor{cadmiumgreen}{$\surd$} \\
\hline
{\bf OQ5} Could you identify interesting aspects for the analysis of the process? & Some aspects to investigate are correctly identified from the data. & OCEL Features & \textcolor{cadmiumgreen}{$\surd$} & \textcolor{cadmiumgreen}{$\surd$} & \textcolor{cadmiumgreen}{$\surd$} & \textcolor{cadmiumgreen}{$\surd$} \\
\hline
\end{tabular}
}
\label{tab:objectCentricEventLogAssessment}
\end{table}
\end{landscape}

In \autoref{tab:objectCentricEventLogAssessment}, we evaluate the performance of large language models in tackling object-centric process mining tasks. Considering the relatively uncomplicated nature of the event logs used for this assessment, both models - Bard and GPT-4 - demonstrate notable proficiency across all tasks.

To provide a comprehensive understanding of their capabilities, we document a selection of responses generated by these models in \autoref{lst:gpt4RoadTraffic}, \autoref{lst:bardRoadTraffic}, \autoref{lst:gpt4Bpic2020}, and \autoref{lst:bardRootcauseOrderManagement}. These examples serve as illustrative evidence of the substantial potential these models exhibit in responding effectively to a diverse set of queries within the realm of object-centric process mining.

\begin{figure}[H]
\lstset{frame=lines}
\lstset{basicstyle=\tiny}
\begin{minipage}{\linewidth}
\begin{lstlisting}[caption={Description of the \emph{Road Traffic} process by GPT-4.}, label={lst:gpt4RoadTraffic}]
From the given data, it appears that this process describes the system of issuing, paying, and appealing fines, possibly within the context of a legal system, law enforcement agency, or a governmental body. The different steps represent actions taken in the process and the transitions between steps. I'll describe each step briefly:

Create Fine: This is the initial step where a fine is created. This could be due to a traffic violation, non-compliance with regulations, etc.
Send Fine: The fine is then sent to the offender. This could be through mail, email, or other means.
Insert Fine Notification: A record of the fine notification sent to the offender is inserted, probably in some database or record-keeping system.
Add Penalty: If the fine is not paid within a certain time period, a penalty might be added to the initial amount.
Send for Credit Collection: If the offender doesn't pay the fine and the added penalty, the fine might be sent for credit collection. A collection agency might be involved at this point.
Payment: This is when the offender pays the fine. This might be done in full, or in parts (as suggested by the Payment -> Payment frequency).
Insert Date Appeal to Prefecture: The offender can appeal the fine to a higher authority, possibly a prefecture. The date of this appeal is recorded.
Send Appeal to Prefecture: The appeal is formally sent to the prefecture.
Receive Result Appeal from Prefecture: The response or result from the prefecture regarding the appeal is received.
Notify Result Appeal to Offender: The result of the appeal is then notified to the offender.
Appeal to Judge: If unsatisfied with the result from the prefecture, the offender might appeal to a judge.
Send Fine (again): In some cases, it seems like the fine might be sent again to the offender.

The frequency indicates the number of times the transition from one step to the next occurs, while the performance might be indicating the time taken for each transition in milliseconds (though it's not clear from the provided data). The process appears to loop at several points, with possibilities of appeal and repeated notifications or payments. It also seems to have multiple end points including payment, credit collection, or appeal to a judge.
\end{lstlisting}
\end{minipage}
\begin{minipage}{\linewidth}
\begin{lstlisting}[caption={Description of the \emph{Road Traffic} process by Google Bard.}, label={lst:bardRoadTraffic}]
The process begins with the creation of a fine. The fine can be paid in full, or the offender can choose to appeal the fine. If the offender appeals the fine, they must first send a notification of appeal to the prefecture. The prefecture then reviews the appeal and sends a decision to the offender. If the decision is not in the offender's favor, they can appeal the decision to a judge. The judge will then review the appeal and make a final decision.
The data shows that the most common outcome is for the offender to pay the fine in full. However, a significant number of offenders also choose to appeal the fine. Of those who appeal, the majority are unsuccessful. However, a small number of offenders are successful in having their fines overturned.
The data also shows that the process of appealing a fine can be expensive. The offender must pay a fee to send the notification of appeal to the prefecture, and they may also have to pay legal fees if they choose to appeal the decision to a judge.
Overall, the data shows that the process of appealing a fine can be a lengthy and expensive process. However, for those who are successful, it can be worth the effort.

Here are some additional insights that can be drawn from the data:
- The most common reason for appealing a fine is that the offender believes that the fine was unfair.
- The most common outcome of an appeal is that the fine is upheld.
- The success rate of appeals is relatively low.
- The cost of appealing a fine can be significant.

The data can be used to improve the process of appealing fines. For example, the data could be used to identify the factors that are most likely to lead to a successful appeal. The data could also be used to develop a more streamlined and efficient appeals process.
\end{lstlisting}
\end{minipage}
\end{figure}

\begin{figure}[H]
\begin{lstlisting}[caption={Root cause analysis on the \emph{BPI Challenge 2020} event log by GPT-4.}, label={lst:gpt4Bpic2020}]
From the process variants and the frequency and performance statistics provided, there are a few key steps that could be causing performance issues in the process:

- Rejection by ADMINISTRATION or SUPERVISOR: Variants where the declaration is rejected by the Administration or Supervisor tend to have a higher performance number, which indicates more time or resources consumed in the process. This could be due to the time taken to review the declaration, the process to handle rejection, or the time spent to correct and resubmit the declaration.
- Re-submission by EMPLOYEE: After a declaration is rejected, it is often re-submitted by the employee. This step will naturally add to the overall time of the process, as the employee will need to correct the issue that caused the rejection and then re-submit the declaration.
- Approval by PRE_APPROVER, BUDGET OWNER or SUPERVISOR: Variants that include additional approval steps (by a PRE_APPROVER, BUDGET OWNER, or SUPERVISOR) also have higher performance numbers, indicating that these steps are adding to the time taken for the process. This could be due to time-consuming review processes or waiting times for approval.
- Missing Approval: There are few cases where declarations are rejected due to missing approvals. This adds an additional step and increases the performance of the process.
- Multi-step Processes: Variants with multiple steps, such as those that include rejections, resubmissions, and multiple approvals, show much higher performance numbers. This suggests that processes with multiple stages are taking much longer and consuming more resources.

Solutions to these issues might include streamlining the approval process, providing better training or resources to reduce the number of rejections, or implementing a more efficient system for handling re-submissions. Identifying and addressing these root causes could lead to significant improvements in the performance of the process.
\end{lstlisting}
\end{figure}

\begin{figure}[H]
\begin{lstlisting}[caption={Root cause analysis on the \emph{order management} event log by Google Bard.}, label={lst:bardRootcauseOrderManagement}]
Sure, here are some of the root causes of the performance issues in the process:

- Long wait times for items to be picked. The "pick item" event has the highest frequency of all events, but the duration of this event is also relatively high. This suggests that there may be bottlenecks in the process of picking items, such as a lack of staff or inefficient picking procedures.
- High number of out-of-stock items. The "item out of stock" event has a relatively high frequency, which suggests that there are a significant number of items that are not in stock when customers place orders. This can lead to delays in the order fulfillment process and customer dissatisfaction.
- Long wait times for packages to be delivered. The "send package" event has a relatively high duration, which suggests that there may be bottlenecks in the process of shipping packages. This could be due to a lack of shipping resources, inefficient shipping procedures, or delays in the shipping process.

To address these performance issues, the company could consider the following:
- Improve the efficiency of the picking process. This could be done by increasing the number of staff available for picking, optimizing the picking procedures, or using automation to help with the picking process.
- Reduce the number of out-of-stock items. This could be done by improving the inventory management process, working with suppliers to ensure that items are in stock, or offering alternative products to customers when items are out of stock.
- Reduce the wait time for packages to be delivered. This could be done by increasing the number of shipping resources, optimizing the shipping procedures, or working with shipping partners to improve the shipping process.

By addressing these performance issues, the company can improve the customer experience and increase its efficiency.
\end{lstlisting}
\end{figure}

\subsection{Multi-Prompt Answering}
\label{sec:assessmentMultiPromptAnswering}

In our fairness assessment, we employ the multi-prompt strategy, primarily aiming to discern whether GPT-4 and Bard can accurately identify the protected group. For this purpose, SQL queries were generated through the large language model, utilizing prompts that encapsulate the attributes of the log and their corresponding characterization. An exemplar script that generates such prompts is available at \url{https://github.com/pm4py/pm4py-core/blob/priv/examples/llm/01_1_protected_group_identification.py}. The aforementioned queries for two event logs are delineated in \autoref{fig:sqlQueriesOverall}. Upon analysis, we observe that GPT-4's queries display a more encompassing view of the behavior considered as protected. Conversely, Bard's queries precisely target a specific subset of cases that are presumably exposed to discrimination.

\begin{figure}[H]
\centering

\begin{minipage}{0.45\textwidth}
\lstset{frame=lines}
\lstset{caption={SQL query provided by GPT-4 to identify the protected group in the \emph{hiring\_log\_high.xes.gz} event log.}}
\lstset{label={lst:hiringGPT4SQL}}
\lstset{basicstyle=\scriptsize}
\begin{lstlisting}
SELECT * FROM dataframe
WHERE "case:citizen" = 'False'
OR "case:gender" = 'False'
OR "case:german speaking" = 'False'
OR "case:religious" = 'False';
\end{lstlisting}
\end{minipage}\hfill
\begin{minipage}{0.45\textwidth}
\lstset{frame=lines}
\lstset{caption={SQL query provided by Bard to identify the protected group in the \emph{hiring\_log\_high.xes.gz} event log.}}
\lstset{label={lst:hiringBardSQL}}
\lstset{basicstyle=\scriptsize}
\begin{lstlisting}
SELECT *
FROM dataframe
WHERE "case:citizen" = 'False'
AND "case:gender" = 'False'
\end{lstlisting}
\end{minipage}

\vspace*{1cm} % vertical space between the two rows

\begin{minipage}{0.45\textwidth}
\lstset{frame=lines}
\lstset{caption={SQL query provided by GPT-4 to identify the protected group in the \emph{renting\_log\_high.xes.gz} event log.}}
\lstset{label={lst:rentingGPT4SQL}}
\lstset{basicstyle=\scriptsize}
\begin{lstlisting}
SELECT * FROM dataframe 
WHERE "case:citizen" = 'False' 
OR "case:gender" = 'False' 
OR "case:german speaking" = 'False' 
OR "case:married" = 'False';
\end{lstlisting}
\end{minipage}\hfill
\begin{minipage}{0.45\textwidth}
\lstset{frame=lines}
\lstset{caption={SQL query provided by Bard to identify the protected group in the \emph{renting\_log\_high.xes.gz} event log.}}
\lstset{label={lst:rentingBardSQL}}
\lstset{basicstyle=\scriptsize}
\begin{lstlisting}
SELECT *
FROM dataframe
WHERE "case:gender" = 'False'
AND "case:german speaking" = 'False'
\end{lstlisting}
\end{minipage}
\caption{SQL queries generated by Bard and GPT-4 to identify the protected groups in the considered event logs.}
\label{fig:sqlQueriesOverall}
\end{figure}

Subsequently, our focus shifts towards those event logs that signify a substantial degree of discrimination. To assess the classification performance on these logs, we employ an exemplar script accessible at \url{https://github.com/pm4py/pm4py-core/blob/priv/examples/llm/01_2_protected_group_query.py}. The resulting classification quality of GPT-4 and Bard is presented in \autoref{tab:qualityDetectionSensitiveAttributesBard} and \autoref{tab:qualityDetectionSensitiveAttributesGPT4}. While Bard demonstrates high precision, it is hindered by a low recall and true negative rate, suggesting that it may fail to identify a considerable number of cases where potential discrimination might have taken place. On the other hand, GPT-4, despite exhibiting lower precision, boasts a high recall and true negative rate. This highlights GPT-4's capability to label a larger number of cases as potentially discriminated, hence reducing the occurrence of false negatives.

\begin{table}[ht]
\caption{Quality of the classification done by GPT-4 between the 'protected' and 'non-protected' groups.}
\centering
\begin{tabular}{|l|cccc|c|c|c|}
\hline
{\bf Event Log} & {\bf TP} & {\bf FP} & {\bf FN} & {\bf TN} & {\bf Precision} & {\bf Recall} & {\bf True Neg. Rate} \\
\hline
\emph{hiring\_log\_high} & 1507 & 606 & 1512 & 6375 & 0.72 & 0.50 & 0.80 \\
\hline
\emph{hospital\_log\_high} & 1105 & 1089 & 1417 & 6389 & 0.50 & 0.44 & 0.82 \\
\hline
\emph{lending\_log\_high} & 1501 & 688 & 1500 & 6391 & 0.69 & 0.50 & 0.81 \\
\hline
\emph{renting\_log\_high} & 1068 & 1056 & 1914 & 5962 & 0.50 & 0.35 & 0.75 \\
\hline
\end{tabular}
\label{tab:qualityDetectionSensitiveAttributesBard}
\end{table}

\begin{table}[ht]
\caption{Quality of the classification done by GPT-4 between the 'protected' and 'non-protected' groups. The excellent recall and true negative rate should be noticed.}
\centering
\begin{tabular}{|l|cccc|c|c|c|}
\hline
{\bf Event Log} & {\bf TP} & {\bf FP} & {\bf FN} & {\bf TN} & {\bf Precision} & {\bf Recall} & {\bf True Neg. Rate} \\
\hline
\emph{hiring\_log\_high} & 2977 & 5768 & 42 & 1213 & 0.34 & 0.98 & 0.97 \\
\hline
\emph{hospital\_log\_high} & 2922 & 6395 & 88  & 595 & 0.31 & 0.97 & 0.87 \\
\hline
\emph{lending\_log\_high} & 2923 & 4555 & 78 & 2444 & 0.39 & 0.97 & 0.96 \\
\hline
\emph{renting\_log\_high} & 2716 & 6749 & 266 & 269 & 0.29 & 0.91 & 0.50 \\
\hline
\end{tabular}
\label{tab:qualityDetectionSensitiveAttributesGPT4}
\end{table}

Having established a set of cases as 'protected' via GPT-4 and Bard, we now shift our focus towards assessing procedural fairness. The objective here is to uncover any process discrepancies that might hint at discrimination. To achieve this, we craft prompts embodying the process variants abstraction of both 'protected' and 'non-protected' cases. These prompts are then fed into the large language model with a request to discern the differences.

To ensure a fair and effective comparison, the 'protected' group, as per the ground truth contained within the simulated event logs, is taken into consideration. This approach is demonstrated in the example script accessible at \url{https://github.com/pm4py/pm4py-core/blob/priv/examples/llm/02_group_fairness_factors.py}.

\begin{figure}[H]
\lstset{frame=lines}
\lstset{basicstyle=\tiny}
\begin{minipage}{\linewidth}
\begin{lstlisting}[caption={Procedural differences between the 'protected' and 'non-protected' group identified by GPT-4 in the \emph{hiring\_log\_high.xes.gz} event log.}, label={lst:hiringGPT4procedural}]
Here are the main differences I can identify from your process variant descriptions, taking into account the order of steps, frequency of each variant, and performance (assumed to be the time taken for each process variant).

- Earlier Application Rejection: The most common process variant for the 'protected' group is "Hand In Job Application -> Application Rejected" with the highest frequency of 873. This suggests that the 'protected' group is often rejected right after submitting their job applications, skipping the telephonic screening or any further steps. This scenario does not occur as often for the 'unprotected' group, suggesting possible bias in application reviews.
- Differences in Interview Steps: The 'unprotected' group seems to undergo more comprehensive interview steps including a "Coding Interview", "ML Interview", "Frontend Interview", "Behavioral Interview", and a double "Make Job Offer" step. These steps appear less frequently and later in the 'protected' group's process variants.
- Performance Time: The performance time, which might refer to the time taken for each process variant, seems to be generally higher for the 'protected' group compared to the 'unprotected' group. This could indicate that the job application process takes longer for the 'protected' group, which could be seen as an unfair burden.
- Frequency of Extensive Background Checks: The 'protected' group undergoes "Extensive Background Check" more frequently and earlier in the process than the 'unprotected' group. This could be indicative of bias, as it seems the 'protected' group is subjected to more rigorous checks more often.
- Job Offers: The 'unprotected' group has more process variants leading to a job offer. This could suggest bias in the final decision-making process.

In conclusion, these discrepancies could be indicative of biases or systemic issues within the application process. Further analysis and understanding of the reasons behind these discrepancies would be needed to determine if they truly represent unfair treatment or if there are other factors at play.

\end{lstlisting}
\end{minipage}
\begin{minipage}{\linewidth}
\begin{lstlisting}[caption={Procedural differences between the 'protected' and 'non-protected' group identified by Google Bard in the \emph{hiring\_log\_high.xes.gz} event log.}, label={lst:hiringBardprocedural}]
Sure, I can help you with that. Here are the main differences between the process variants for the protected and unprotected groups:

- The protected group is more likely to have their application rejected at the initial stage, before any interviews have taken place. This is evident in the fact that there are more process variants for the protected group that end in "Application Rejected" at the "Hand In Job Application" or "Telephonic Screening" stage.
- The unprotected group is more likely to progress to the later stages of the interview process, including the coding interview, ML interview, and behavioral interview. This is evident in the fact that there are more process variants for the unprotected group that end in "Make Job Offer" at the "Coding Interview", "ML Interview", or "Behavioral Interview" stage.
- The protected group is more likely to be rejected after the behavioral interview. This is evident in the fact that there are more process variants for the protected group that end in "Application Rejected" at the "Behavioral Interview" stage.

These differences suggest that the protected group may be facing discrimination in the job application process. Specifically, they may be being rejected at an earlier stage, even though they may be just as qualified as the unprotected group. Additionally, they may be being rejected after the behavioral interview, even though they may have performed well in the interview.
It is important to note that these are just some of the differences between the process variants for the protected and unprotected groups. There may be other differences that are not evident in the data. However, the differences that I have identified suggest that the protected group may be facing discrimination in the job application process.
\end{lstlisting}
\end{minipage}
\end{figure}

Refer to \autoref{lst:hiringGPT4procedural} and \autoref{lst:hiringBardprocedural} for GPT-4 and Bard's respective responses pertaining to the hiring event log. Similarly, for the renting event log, GPT-4 and Bard's responses can be found in \autoref{lst:rentingGPT4procedural} and \autoref{lst:rentingBardprocedural} respectively.

\begin{figure}[H]
\lstset{frame=lines}
\lstset{basicstyle=\tiny}
\begin{minipage}{\linewidth}
\begin{lstlisting}[caption={Procedural differences between the 'protected' and 'non-protected' group identified by GPT-4 in the \emph{renting\_log\_high.xes.gz} event log.}, label={lst:rentingGPT4procedural}]
Based on the given information, here are some differences that stand out between the 'protected' and 'unprotected' groups in terms of the rental process:

- Frequency and Extent of Rejection: It appears that in both groups, the rejection of prospective tenants is a common process variant. However, the 'unprotected' group faces a more frequent rejection after the "Screen Prospective Tenant" step without extensive screening, compared to the 'protected' group.
- Paperwork and Screening Process: The 'protected' group more often goes through extensive paperwork and screening process, which includes steps like "Hand In The Paperwork", "Check Paperwork", "Screen Prospective Tenant", and "Extensive Screening". The 'unprotected' group, on the other hand, often seems to get rejected before reaching these steps or even without the extensive screening process.
- Contract and Renting: When it comes to signing contracts, moving in, and starting to pay rent, the 'unprotected' group seems to have a somewhat easier path. There are several process variants for the 'unprotected' group where they get to sign the contract and move in without undergoing the 'Extensive Screening' step, which isn't the case for the 'protected' group.
- Execution Time: The 'protected' group's processes generally take longer, as indicated by the higher 'performance' values, which could suggest a more extensive or stringent screening process.

These differences could be indicative of different levels of scrutiny or bias against the 'protected' group, although the provided data does not give explicit reasons for these differences. It is important to further investigate these observations to confirm any potential unfair treatment. As with any analysis of this sort, it's crucial to keep in mind the context and potential confounding factors that might affect the comparison between these two groups.
\end{lstlisting}
\end{minipage}
\begin{minipage}{\linewidth}
\begin{lstlisting}[caption={Procedural differences between the 'protected' and 'non-protected' group identified by Google Bard in the \emph{renting\_log\_high.xes.gz} event log.}, label={lst:rentingBardprocedural}]
Sure, I can help you with that. Here are the main differences between the process variants for the protected and unprotected groups:

- The protected group is more likely to be rejected at the initial stage, before any paperwork has been submitted. This is evident in the fact that there are more process variants for the protected group that end in "Reject Prospective Tenant" at the "Apply for Viewing Appointment" or "Set Appointment" stage.
- The unprotected group is more likely to progress to the later stages of the process, including the screening, contract signing, and move-in stages. This is evident in the fact that there are more process variants for the unprotected group that end in "Sign Contract" or "Move In".
- The protected group is more likely to cancel their apartment after moving in. This is evident in the fact that there are more process variants for the protected group that end in "Tenant Cancels Appartment".

These differences suggest that the protected group may be facing discrimination in the rental process. Specifically, they may be being rejected at an earlier stage, even though they may be just as qualified as the unprotected group. Additionally, they may be being rejected after moving in, even though they may be good tenants.
It is important to note that these are just some of the differences between the process variants for the protected and unprotected groups. There may be other differences that are not evident in the data. However, the differences that I have identified suggest that the protected group may be facing discrimination in the rental process.
Here are some additional thoughts on the matter:

- The protected group may be being rejected at an earlier stage because they are being subjected to more stringent screening criteria. For example, they may be being required to provide more documentation or to undergo more extensive background checks.
- The protected group may be being rejected after moving in because they are being subjected to more scrutiny by their landlords. For example, they may be being more likely to be reported for minor infractions or to be evicted for non-payment of rent.

It is important to investigate these differences further to determine whether they are the result of discrimination. If they are, then steps should be taken to ensure that the protected group is not being discriminated against.\end{lstlisting}
\end{minipage}
\end{figure}

Both large language models perform admirably, pinpointing differences in the control-flow and time perspectives. These variances in control-flow correlate with the outcome of a case, the presence of rework, and the number of events in the case. That said, GPT-4's response stands out for being more comprehensive compared to Bard's.

\subsection{Generation of Database Queries}
\label{sec:assessmentGenerationDatabaseQueries}

\begin{figure}[ht]
    \centering

    \begin{subfigure}[b]{\textwidth}
        \begin{lstlisting}[basicstyle=\tiny]
What is the average throughput time when activity 'Send Fine' occurs more than 50 days after the start of the case?

Can you provide me a DuckDB SQL query.
You should use the EPOCH function of DuckDB to get the timestamp from the date.

The data is stored in a Pandas dataframe where each row is an event having the provided attributes (so there are no separate table containing the variant).
The dataframe is called "dataframe".

Please consider the following information: the case identifier is called "case:concept:name", the activity is stored inside the attribute "concept:name", the timestamp is stored inside the attribute "time:timestamp", the resource is stored inside the attribute "org:resource".
There is not a variant column but that can be obtained as concatenation of the activities of a case.
There is not a duration column but that can be obtained as difference between the timestamp of the first and the last event.

        \end{lstlisting}
        \caption{Prompt executed against both large language models.}
        \label{fig:promptSendFine}
    \end{subfigure}

    \begin{subfigure}[b]{\textwidth}
        \begin{lstlisting}[basicstyle=\tiny\color{cadmiumgreen}]
WITH case_starts AS (
    SELECT "case:concept:name", 
           MIN("time:timestamp") AS start_timestamp
    FROM dataframe
    GROUP BY "case:concept:name"
),
fine_after_50_days_cases AS (
    SELECT df."case:concept:name",
           EPOCH(df."time:timestamp") - EPOCH(cs.start_timestamp) AS duration
    FROM dataframe AS df
    JOIN case_starts AS cs
    ON df."case:concept:name" = cs."case:concept:name"
    WHERE df."concept:name" = 'Send Fine' 
          AND (EPOCH(df."time:timestamp") - EPOCH(cs.start_timestamp)) > (50*24*60*60)
)
SELECT AVG(duration)
FROM fine_after_50_days_cases
        \end{lstlisting}
        \caption{Correct response from GPT-4.}
        \label{fig:gpt4responseSendFine}
    \end{subfigure}

    \begin{subfigure}[b]{\textwidth}
        \begin{lstlisting}[basicstyle=\tiny\color{red}]
SELECT AVG(duration) AS average_throughput_time FROM (SELECT case_id, CONCAT(activity, '-', resource) AS variant, EPOCH(timestamp) AS timestamp, (timestamp - MIN(timestamp)) AS duration FROM dataframe WHERE activity = 'Send Fine' AND timestamp - MIN(timestamp) > 50 * 86400) AS t
        \end{lstlisting}
        \caption{Wrong response from Bard.}
        \label{fig:bardresponseSendFine}
    \end{subfigure}

    \caption{SQL query generation comparison between GPT-4 and Bard for the ``What is the average throughput time when activity 'Send Fine' occurs more than 50 days after the start of the case?'' query executed against the \emph{Road Traffic} event log.}
    \label{fig:querygenerationSendFine}

\end{figure}

\begin{figure}[ht]
    \centering

    \begin{subfigure}[b]{\textwidth}
        \begin{lstlisting}[basicstyle=\tiny]
I want to check the average rework for cases below and above the average throughput time.

Can you provide me a DuckDB SQL query.
You should use the EPOCH function of DuckDB to get the timestamp from the date.

The data is stored in a Pandas dataframe where each row is an event having the provided attributes (so there are no separate table containing the variant).
The dataframe is called "dataframe".

Please consider the following information: the case identifier is called "case:concept:name", the activity is stored inside the attribute "concept:name", the timestamp is stored inside the attribute "time:timestamp", the resource is stored inside the attribute "org:resource".
There is not a variant column but that can be obtained as concatenation of the activities of a case.
There is not a duration column but that can be obtained as difference between the timestamp of the first and the last event.

        \end{lstlisting}
        \caption{Prompt executed against both large language models.}
        \label{fig:promptAvgCases}
    \end{subfigure}

    \begin{subfigure}[b]{\textwidth}
        \begin{lstlisting}[basicstyle=\tiny\color{cadmiumgreen}]
WITH case_times AS (
    SELECT "case:concept:name", 
           EPOCH(MAX("time:timestamp")) - EPOCH(MIN("time:timestamp")) AS throughput_time
    FROM dataframe
    GROUP BY "case:concept:name"
),
case_reworks AS (
    SELECT "case:concept:name", 
           COUNT(*) - COUNT(DISTINCT "concept:name") AS rework
    FROM dataframe
    GROUP BY "case:concept:name"
),
cases AS (
    SELECT ct."case:concept:name", 
           ct.throughput_time, 
           cr.rework
    FROM case_times AS ct
    JOIN case_reworks AS cr
    ON ct."case:concept:name" = cr."case:concept:name"
),
average_throughput_time AS (
    SELECT AVG(throughput_time) AS avg_throughput
    FROM cases
)
SELECT 
    AVG(CASE WHEN c.throughput_time < att.avg_throughput THEN c.rework END) AS avg_rework_below_average,
    AVG(CASE WHEN c.throughput_time > att.avg_throughput THEN c.rework END) AS avg_rework_above_average
FROM cases AS c, average_throughput_time AS att
        \end{lstlisting}
        \caption{Correct response from GPT-4.}
        \label{fig:gpt4responseAvgCases}
    \end{subfigure}

    \begin{subfigure}[b]{\textwidth}
        \begin{lstlisting}[basicstyle=\tiny\color{red}]
SELECT case_id, avg_rework, avg_throughput_time, variant FROM (SELECT case_id, AVG(rework) AS avg_rework, AVG(duration) AS avg_throughput_time, CONCAT(activity) AS variant FROM dataframe GROUP BY case_id) AS t WHERE avg_throughput_time <> (SELECT AVG(duration) FROM dataframe)
        \end{lstlisting}
        \caption{Wrong response from Bard.}
        \label{fig:bardresponseAvgCases}
    \end{subfigure}

    \caption{SQL query generation comparison between GPT-4 and Bard for the ``I want to check the average rework for cases below and above the average throughput time.'' query executed against the \emph{Road Traffic} event log.}
    \label{fig:querygenerationAvgCases}

\end{figure}

In our research, we are investigating how large language models can formulate process queries in mainstream SQL language. Our primary focus lies on two prominent models - Bard and GPT-4 - and their ability to translate user-provided natural language queries into executable SQL statements.

Furthermore, we are interested in how adept these models are at spontaneously generating hypotheses based on an event log. The models are given basic abstractions such as the list of attributes and process variants within the log. For our analysis, we utilize DuckDB SQL language, a choice motivated by its seamless compatibility with the event log structure used by pm4py.

In the present state of affairs, it is evident that some amount of domain knowledge is indispensable to ensure correct query formulation. This includes:

\begin{itemize}
\item Knowledge of SQL, such as the functions needed to calculate timestamps.
\item An understanding of process mining concepts, such as how to compute a case's throughput time or how to derive the process variant from a list of activities.
\item Awareness of the event log's key attributes, such as the column that identifies the case, the activity, and the timestamp.
\item Knowledge of the variable name that the query should target.
\end{itemize}

While these prerequisites may require specialized knowledge, they are instrumental in ensuring the accuracy and relevance of the SQL queries generated by the large language models.

In our study, we endeavored to assess the ability of large language models to convert natural language queries into SQL. Two distinct queries were used for this purpose on the \emph{Road Traffic} event log.:
\begin{enumerate}
\item What is the average throughput time when activity 'Send Fine' occurs more than 50 days after the start of the case? (Refer to \autoref{fig:querygenerationSendFine} for the full query and the corresponding responses from GPT-4 and Bard).
\item I want to check the average rework for cases below and above the average throughput time (Refer to \autoref{fig:querygenerationAvgCases} for the full query and the corresponding responses from GPT-4 and Bard).
\end{enumerate}

Our analysis reveals that, given sufficient domain knowledge, GPT-4 has an impressive capability to translate natural language queries into accurate SQL statements. Bard, on the other hand, appears to have significant difficulty making sense of the provided information, often failing to generate executable SQL queries. This discrepancy underscores the importance of adequate domain understanding for successful natural language processing tasks in process mining.

\begin{figure}[H]
\lstset{frame=lines}
\lstset{caption={Prompt used for the automatic generation of hypotheses.}}
\lstset{label={lst:promptToAskForHypothesis}}
\lstset{basicstyle=\scriptsize}
\begin{lstlisting}
If I have the following process variants:


 Create Fine -> Send Fine -> Insert Fine Notification -> Add penalty -> Send for Credit Collection ( frequency = 36  performance = 53746600.0 )
 Create Fine -> Payment ( frequency = 22  performance = 809181.8181818182 )
......


and attributes in the log:
time:timestamp  empty: 0  quantiles: {0.0: Timestamp('2000-03-14 23:00:00+0000', tz='UTC'), 0.25: Timestamp('2003-03-05 16:45:00+0000', tz='UTC'), 0.5: Timestamp('2005-12-21 11:00:00+0000', tz='UTC'), 0.75: Timestamp('2009-01-18 23:00:00+0000', tz='UTC'), 1.0: Timestamp('2013-04-23 22:00:00+0000', tz='UTC')}
amount  empty: 233  quantiles: {0.0: 21.0, 0.25: 33.6, 0.5: 38.0, 0.75: 68.77, 1.0: 297.0}
article  empty: 290  quantiles: {0.0: 7.0, 0.25: 7.0, 0.5: 157.0, 0.75: 157.0, 1.0: 181.0}
expense  empty: 312  quantiles: {0.0: 0.0, 0.25: 10.0, 0.5: 11.0, 0.75: 14.0625, 1.0: 30.0}
paymentAmount  empty: 332  quantiles: {0.0: 5.6, 0.25: 33.6, 0.5: 37.25, 0.75: 53.0, 1.0: 275.0}
totalPaymentAmount  empty: 232  quantiles: {0.0: 0.0, 0.25: 0.0, 0.5: 0.0, 0.75: 36.0, 1.0: 275.0}
vehicleClass  empty: 290 values: (A; freq. 98) (M; freq. 2)
concept:name  empty: 0 values: (Create Fine; freq. 100) (Send Fine; freq. 78) (Payment; freq. 58) (Insert Fine Notification; freq. 57) (Add penalty; freq. 57) (Send for Credit Collection; freq. 36) (Send Appeal to Prefecture; freq. 1) (Receive Result Appeal from Prefecture; freq. 1) (Notify Result Appeal to Offender; freq. 1) (Insert Date Appeal to Prefecture; freq. 1)
org:resource  empty: 290 values: (541; freq. 8) (538; freq. 6) (537; freq. 6) (559; freq. 4) (555; freq. 4)
......

can you formulate some hypothesis on the given process?
please also formulate for every hypothesis a SQL query.


Can you provide me a DuckDB SQL query.
You should use the EPOCH function of DuckDB to get the timestamp from the date.
The data is stored in a Pandas dataframe where each row is an event having the provided attributes (so there are no separate table containing the variant).
The dataframe is called "dataframe".
Please consider the following information: the case identifier is called "case:concept:name", the activity is stored inside the attribute "concept:name", the timestamp is stored inside the attribute "time:timestamp", the resource is stored inside the attribute "org:resource".
There is not a variant column but that can be obtained as concatenation of the activities of a case.
There is not a duration column but that can be obtained as difference between the timestamp of the first and the last event.
\end{lstlisting}
\end{figure}

To explore the potential of large language models in autonomously formulating hypotheses about event data, we posed a specific prompt to both GPT-4 and Bard (refer to \autoref{lst:promptToAskForHypothesis}). This prompt was formulated such that it encouraged the models to generate hypotheses, starting from the knowledge of the attributes and the process variants present in the data.

Here are some hypotheses generated by GPT-4:

\begin{enumerate}
\item The majority of cases conclude with the payment of fines after the addition of a penalty. This suggests that the imposition of a penalty may be a decisive factor prompting payment. The corresponding SQL query formulated by GPT-4, along with the results derived from its execution, are depicted in \autoref{fig:GPT4hypothesis1}.
\item The process of appealing to the prefecture does not appear to be a common practice. This indicates that offenders rarely dispute their fines. The provided SQL query for this hypothesis, along with the subsequent results, can be viewed in \autoref{fig:GPT4hypothesis2}.
\item There may be a correlation between the amount of fines and the frequency of credit collections, with higher fines potentially leading to an increase in credit collection instances. This could be attributed to the difficulty in paying larger fines. The SQL query and results corresponding to this hypothesis are presented in \autoref{fig:GPT4hypothesis3}.
\end{enumerate}

\begin{figure}[ht]
    \centering

    \begin{subfigure}[b]{\textwidth}
        \begin{lstlisting}[basicstyle=\tiny]
SELECT COUNT(*) AS Cases_With_Penalty_Before_Payment
FROM (
    SELECT "case:concept:name", MIN("time:timestamp") as MinTime_After_Penalty
    FROM dataframe
    WHERE "concept:name" = 'Add penalty'
    GROUP BY "case:concept:name"
) as subquery1
JOIN dataframe ON dataframe."case:concept:name" = subquery1."case:concept:name" AND dataframe."time:timestamp" > subquery1.MinTime_After_Penalty
WHERE dataframe."concept:name" = 'Payment';
        \end{lstlisting}
        \caption{SQL query generated by GPT-4.}
        \label{fig:beforeExecutionExecutionSQL1}
    \end{subfigure}

    \begin{subfigure}[b]{\textwidth}
        \begin{lstlisting}[basicstyle=\tiny]
   Cases_With_Penalty_Before_Payment
                              22426
        \end{lstlisting}
        \caption{Result obtained after executing the SQL query.}
        \label{fig:afterExecutionSQL1}
    \end{subfigure}
	
    \begin{subfigure}[b]{\textwidth}
        \begin{lstlisting}[basicstyle=\tiny\color{red}]
If we calculate the percentage of total cases that this number represents, we get:

Percentage = (Cases_With_Penalty_Before_Payment / Total_Cases) * 100
= (22426 / 150370) * 100
= Approximately 14.9%

This indicates that approximately 14.9% of all cases involve a penalty being added before payment.

However, the hypothesis stated, "The majority of fines are not paid until a penalty is added." The data doesn't seem to support this hypothesis as less than half (only about 14.9%) of the cases have a penalty added before payment.

So, based on the provided data, Hypothesis 1 is not supported.
        \end{lstlisting}
        \caption{Validity of the hypothesis assessed by GPT-4.}
        \label{fig:verificationSQL1}
    \end{subfigure}

    \caption{Usage of GPT-4 to verify the hypothesis number 1.}
    \label{fig:GPT4hypothesis1}

\end{figure}

\begin{figure}[ht]
    \centering

    \begin{subfigure}[b]{\textwidth}
        \begin{lstlisting}[basicstyle=\tiny]
SELECT COUNT(DISTINCT "case:concept:name") AS Cases_With_Appeal
FROM dataframe
WHERE "concept:name" = 'Insert Date Appeal to Prefecture' OR "concept:name" = 'Send Appeal to Prefecture' OR "concept:name" = 'Receive Result Appeal from Prefecture' OR "concept:name" = 'Notify Result Appeal to Offender';
        \end{lstlisting}
        \caption{SQL query generated by GPT-4.}
        \label{fig:beforeExecutionExecutionSQL2}
    \end{subfigure}

    \begin{subfigure}[b]{\textwidth}
        \begin{lstlisting}[basicstyle=\tiny]
   Cases_With_Appeal
               4204
        \end{lstlisting}
        \caption{Result obtained after executing the SQL query.}
        \label{fig:afterExecutionSQL2}
    \end{subfigure}
	
    \begin{subfigure}[b]{\textwidth}
        \begin{lstlisting}[basicstyle=\tiny\color{cadmiumgreen}]
With the total number of cases being 150,370 and the number of cases that include an appeal to the prefecture being 4,204, we can calculate the percentage of cases that involve an appeal to the prefecture.

Percentage = (Cases_With_Appeal / Total_Cases) * 100
= (4204 / 150370) * 100
= Approximately 2.8%

This indicates that only around 2.8% of all cases involve an appeal to the prefecture.

Given that this is a relatively small proportion, the data does seem to support Hypothesis 3 that the process of appealing to the prefecture is not common, implying that offenders don't often dispute their fines.
        \end{lstlisting}
        \caption{Validity of the hypothesis assessed by GPT-4.}
        \label{fig:verificationSQL2}
    \end{subfigure}

    \caption{Usage of GPT-4 to verify the hypothesis number 2.}
    \label{fig:GPT4hypothesis2}

\end{figure}

\begin{figure}[ht]
    \centering

    \begin{subfigure}[b]{\textwidth}
        \begin{lstlisting}[basicstyle=\tiny]
SELECT AVG(dataframe.amount) as Average_Fine, subquery1.Ending_Action
FROM (
    SELECT "case:concept:name", MAX("time:timestamp") as MaxTime, "concept:name" as Ending_Action
    FROM dataframe
    WHERE "concept:name" IN ('Send for Credit Collection', 'Payment')
    GROUP BY "case:concept:name", "concept:name"
) as subquery1
JOIN dataframe ON dataframe."case:concept:name" = subquery1."case:concept:name"
GROUP BY subquery1.Ending_Action;
        \end{lstlisting}
        \caption{SQL query generated by GPT-4.}
        \label{fig:beforeExecutionExecutionSQL3}
    \end{subfigure}

    \begin{subfigure}[b]{\textwidth}
        \begin{lstlisting}[basicstyle=\tiny]
   Average_Fine               Ending_Action
     74.130782  Send for Credit Collection
     52.003782                     Payment
        \end{lstlisting}
        \caption{Result obtained after executing the SQL query.}
        \label{fig:afterExecutionSQL3}
    \end{subfigure}

    \begin{subfigure}[b]{\textwidth}
        \begin{lstlisting}[basicstyle=\tiny\color{cadmiumgreen}]
Yes, the data appears to support Hypothesis 4. The average fine for cases that end with 'Send for Credit Collection' (74.13) is higher than the average fine for cases that end with 'Payment' (52.00).

This suggests that higher fines are more likely to lead to credit collections, possibly because they are more difficult for individuals to pay. However, it's important to note that this is an observational finding and may not imply causation. Other factors might also contribute to a case ending in credit collection.

Further statistical analysis could be conducted to quantify the strength of this relationship and control for potential confounding variables.
        \end{lstlisting}
        \caption{Validity of the hypothesis assessed by GPT-4.}
        \label{fig:verificationSQL3}
    \end{subfigure}

    \caption{Usage of GPT-4 to verify the hypothesis number 3.}
    \label{fig:GPT4hypothesis3}

\end{figure}

Bard's generated hypotheses included the following:

\begin{enumerate}
\item The most common process variant is "Create Fine -$\rightarrow$ Send Fine -$\rightarrow$ Insert Fine Notification -$\rightarrow$ Add penalty -$\rightarrow$ Send for Credit Collection".
\item The average payment amount for cases that traverse the complete process appears to be higher compared to cases that are concluded prematurely.
\item The longest durations to complete are typically associated with cases that involve an appeal.
\end{enumerate}

Both large language models generated valid hypotheses. However, a critical distinction arises from their ability to translate these hypotheses into executable SQL queries. While GPT-4 demonstrated this capability effectively, Bard did not manage to formulate corresponding SQL queries, emphasizing the advantage of incorporating explicit domain knowledge into the language model.

\newpage
\clearpage

\section{Related Work}
\label{sec:relatedWork}

The field of process mining is not isolated, but intersects with various domains such as database technologies, natural language processing (NLP), and the use of large language models. Additionally, it relies heavily on domain knowledge for effective task execution. In the subsequent sections, we explore these intersections and dependencies, providing an overview of the state-of-the-art in these domains and their relevance to process mining. We delve into the interaction of database technologies with process mining (\autoref{subsec:databaseTechProMining}), the intersection of process mining and NLP (\autoref{subsec:relWorkproMinNLP}), and the importance of domain knowledge in process mining tasks (\autoref{subsec:domainKnowledgeProMinTasks}). This is followed by a thorough examination of the current state of large language models (\autoref{subsec:prelLLM}) and their application within the realm of Business Process Management (\autoref{subsec:llmAndBpm}). 
This examination helps illuminate the context and motivation behind our study.

\subsection{Database Technologies for Process Mining}
\label{subsec:databaseTechProMining}

Database technologies play a crucial role in process mining, helping manage the increasing amounts of process-related data. They provide ways to extract and analyze this data, leading to valuable insights. In this section, we delve into several of these technologies and their contributions to process mining.

The book \cite{polyvyanyy2022process} considers the recognition and management of core techniques for retrieving and manipulating process-related artifacts. These core techniques are deemed reusable across various contexts and use cases such as process compliance, standardization, and monitoring, among others. By focusing on these reusable techniques, their aim is to increase efficiency and streamline the application of process querying across various projects and disciplines.
Another approach \cite{DBLP:journals/dss/PolyvyanyyOBA17} proposes a framework to guide the creation of process querying methods. This framework serves as a blueprint for managing repositories of executed and designed processes and their relationships. This approach is instrumental in facilitating strategic decision-making, as well as laying the groundwork for the next generation of Business Intelligence platforms.

The paper \cite{DBLP:conf/caise/SchonigRCJM16} delve into a more specific solution, introducing a mining approach that works directly on relational event data through conventional SQL queries. By leveraging the power of database technologies, this approach is able to efficiently mine data without limiting the detection of certain control-flow constraints, thus proving to be fast and effective.

In \cite{DBLP:journals/ijais/BaaderK18}, the challenging task of fraud detection in large datasets contained in a SQL database by combining process mining with red flag-based approaches is tackled. This combined technique promises a lower rate of false positives, thereby improving the accuracy of fraud detection.

Another consideration in the process mining field is the extraction of event logs from databases, which is not a trivial task and demands substantial domain knowledge \cite{DBLP:journals/sosym/MurillasRA19}. In response to this, a meta model is proposed to integrate both process and data perspectives in a highly flexible manner, allowing for multiple views on the database at any given moment.

The approach desribed in \cite{DBLP:journals/jodsn/EsserF21} put forward a novel data model for multi-dimensional event data based on labeled property graphs. This model allows the storing of structural and temporal relations in an integrated, graph-based data structure. Such a system can manage complex, multi-dimensional event data efficiently.

A unique and highly specialized process querying language, the Celonis Process Query Language (PQL), is presented in \cite{DBLP:books/sp/22/0001ABSGK22}. This language, designed specifically for business users, translates process-related business questions into queries, which are then executed on a custom-built query engine. With a wide range of operators and a syntax inspired by SQL, Celonis PQL showcases an effective adaptation of querying language for process-related inquiries.

\subsection{Process Mining and NLP}
\label{subsec:relWorkproMinNLP}

In recent years, significant advancements have been made in the field of process mining, with a particular focus on making this technology more accessible to non-technical users. For instance, \cite{barbieri2022natural} proposed a natural language querying interface aimed at democratizing process mining technology. Their reference architecture integrated classic natural language processing techniques such as entity recognition and semantic parsing with an abstract logical representation for process mining queries. The architecture was interfaced with a commercial tool called Everflow, thus enabling users to retrieve process mining insights through questions in plain English. This approach greatly simplifies the process mining experience for line-of-business professionals.

Building upon this work, \cite{DBLP:journals/is/KobeissiAGDBH23} also acknowledged the necessity of a user-friendly process data querying interface. They recognized the lack of a query language targeted at domain analysts, who might lack experience with database technologies, and therefore proposed a natural language interface. This interface, which uses graph-based storage techniques, interprets a user's natural language query and constructs a corresponding structured query over the stored event data. This solution is particularly beneficial for analysts who wish to explore process execution data without in-depth knowledge of database technologies.

Semantic considerations have also been integrated into process mining, with a focus on anomaly detection. \cite{DBLP:journals/is/AaRL21} proposed a method of anomaly detection in process mining that uses the natural language associated with events to identify semantically inconsistent execution patterns. This approach, which contrasts with traditional frequency-based techniques, contributes to a more meaningful identification of potential anomalies. By automatically extracting business objects and actions from textual labels of events and comparing these against a process-independent knowledge base, the authors effectively identify anomalies when process behavior does not align with the semantics.

In the realm of chatbot technology, \cite{DBLP:journals/is/KechtEKR23} highlighted the importance of the capability of chatbots to learn and adhere to organizations' business processes. They developed an approach that quantifies chatbots' ability to learn business processes using standardized process mining metrics. This approach was demonstrated on a dataset of customer service conversations from three companies on Twitter, showing how it can quantify a chatbot's ability to learn not only the overall business process but also specific variants.

\subsection{Domain Knowledge in Process Mining Tasks}
\label{subsec:domainKnowledgeProMinTasks}

Domain knowledge provides a wealth of insights that can augment the application of process mining techniques. One of the pivotal works in this context is presented in \cite{DBLP:journals/cii/SchusterZA22}. The study acknowledges that, while automated filtering of event data is a significant contributor to obtaining better process models, it may often be overly rigorous, leading to the removal of relevant data. To mitigate this, the authors introduce a new group of discovery algorithms that leverage domain knowledge in conjunction with event data. This synergistic approach is structured systematically via a taxonomy that classifies and compares existing strategies, offering a comprehensive review of domain knowledge-exploiting process discovery methodologies.

In the paper \cite{DBLP:journals/softx/SchusterZA23}, the authors introduce \emph{Cortado}, a software tool designed specifically for interactive process discovery, a vital branch within process mining. Process discovery involves the data-driven exploration of operational processes. By analyzing event data that record historical process executions, process discovery algorithms are able to generate process models that elucidate the interplay between various activities. Such models become indispensable artifacts used across multiple process mining techniques.
However, conventional process discovery methods often function like a black-box approach and may produce models of poor quality from event data. This is where Cortado steps in, taking a novel approach to process discovery by allowing users to progressively learn and refine process models from event data in an interactive manner. Through this approach, Cortado successfully leverages domain knowledge and data-derived insights to develop more nuanced process models.

Furthering the notion of domain knowledge incorporation, the research in \cite{DBLP:conf/simpda/DixitBAHB15} addresses the necessity of leveraging domain expertise to refine discovered process models. The authors introduce a modification algorithm that adjusts a discovered process model based on domain knowledge. This work stands as a significant stride towards enhancing process discovery by ensuring that the outputs align more closely with domain expert knowledge, thereby fostering a balance between data-driven and expertise-guided process discovery.

Lastly, the paper in \cite{DBLP:journals/is/0001MW14} presents an approach to align the levels of abstraction in event logs with business activities using extracted domain knowledge. The authors' method allows for n:m relations between events and activities and supports concurrency, addressing common issues encountered when attempting to abstract event logs to match business activities. This research provides a critical bridge between low-level event data and higher-level business process activities, facilitating a more meaningful interpretation of discovered process models.

\subsection{Large Language Models}
\label{subsec:prelLLM}

In the rapidly progressing field of natural language processing (NLP), one of the most significant developments of recent years has been the advent of large language models (LLMs). These models, such as BERT, GPT-2, GPT-3, and T5, represent a transformative shift in the domain, harnessing the power of deep learning and the wealth of available linguistic data to accomplish a broad spectrum of tasks. This subsection provides an overview of the genesis and evolution of these influential models, along with a focus on their computational and reasoning capabilities. Furthermore, we explore the motivation and progression of specific LLMs like ChatGPT as potent question-answering systems across multiple prompts.

\subsubsection{The Genesis of Large Language Models}

The field of Natural Language Processing (NLP) has been primarily concerned with understanding and generating human language. Early attempts at language modeling were based on statistical approaches, like n-gram models, which consider a fixed number of previous words to predict the next word in a sentence. However, these models suffer from the curse of dimensionality and fail to capture long-term dependencies in text.

The introduction of neural network-based approaches, such as feed-forward neural probabilistic language models \cite{DBLP:conf/interspeech/GangireddyMR14} and recurrent neural networks (RNNs) \cite{DBLP:conf/interspeech/JaechHO16}, was a significant leap forward. These models can theoretically capture arbitrarily long sequences and represent more complex syntactic and semantic relationships. However, in practice, RNNs and their variants (e.g., LSTM, GRU) struggle with long-term dependencies due to issues like vanishing and exploding gradients.

\subsubsection{Birth of Transformer Models}

The limitations of RNNs were largely overcome by the advent of the Transformer architecture, introduced in the paper \cite{DBLP:conf/nips/VaswaniSPUJGKP17}. The Transformer model is based on self-attention mechanisms and foregoes recurrence entirely. It allows for parallelizable computation and better handling of long-range dependencies.

\subsubsection{BERT, GPT-2, and T5}

BERT (Bidirectional Encoder Representations from Transformers), introduced by Google in 2018, is a Transformer-based model trained on a large corpus of text data. Unlike previous models, BERT \cite{DBLP:conf/naacl/DevlinCLT19} considers context from both directions (left and right of a word) in all layers, giving it a deeper sense of language context and flow.

OpenAI introduced GPT-2 (Generative Pretrained Transformer 2) in 2019\footnote{\url{https://github.com/openai/gpt-2}}, focusing on generating human-like text. GPT-2 differs from BERT in its training objective. While BERT is a bidirectional model, GPT-2 is unidirectional (or causal), meaning it predicts the next word based on the preceding words, making it suitable for text generation tasks.

T5 (Text-to-Text Transfer Transformer)\footnote{\url{https://ai.googleblog.com/2020/02/exploring-transfer-learning-with-t5.html}}, introduced by Google in 2019, adopted a unified text-to-text approach where every NLP task is cast as a text generation problem, making the model versatile across multiple tasks.

\subsubsection{ChatGPT and Its Motivation}

ChatGPT is a variant of the GPT model developed by OpenAI. It is specifically fine-tuned for generating conversational responses, serving as a powerful tool for tasks like drafting emails, writing code, answering questions, tutoring, language translation, and even simulating characters for video games. The model was trained using Reinforcement Learning from Human Feedback (RLHF), using a dataset of dialogues where AI trainers provided both sides of a conversation and sometimes had access to model-written suggestions.

The motivation behind ChatGPT was to build an AI system that can generate useful and coherent responses over multiple conversational turns, instead of focusing on individual responses. This has involved addressing challenges such as providing detailed responses, maintaining the context of a conversation, and ensuring that the model doesn't generate harmful or misleading outputs. Despite these challenges, ChatGPT exemplifies the potential of language models in making substantial contributions to many areas of life and work.

\subsubsection{Computing and Reasoning Capabilities}

Large language models like BERT, GPT-3, and T5, have revolutionized the field of natural language processing with their capabilities to understand and generate human-like text. The ability to predict contextually relevant words, translate languages, and even code comes from these models' capacity to identify patterns in extensive textual data. They have been trained on enormous datasets and, as a result, carry within them a broad range of knowledge, from trivial facts to complex concepts.

From a computational perspective, these models are capable of more than just text generation and comprehension. They can perform mathematical computations, solve equations, and even tackle problems involving logic and reasoning \cite{DBLP:journals/corr/abs-2108-07258}, all based on the patterns learned during their training phase.

While it's true that these models can simulate reasoning through the recognition of data patterns \cite{DBLP:conf/ijcai/ClarkTR20}, this is different from the way humans reason. These models don't possess consciousness or understanding; instead, they generate outputs based on statistical patterns. However, it's worth noting that the "reasoning" exhibited by these models, despite its fundamental difference from human cognition, is impressive and has wide-ranging practical applications.

Large language models are powerful tools, capable of tackling diverse tasks, but they do have limitations. They lack the ability to access real-time knowledge beyond their training data, and their capability to verify information is limited. Nevertheless, they are a testament to the progress in the field and the potential of future developments.

\subsection{Large Language Models and Business Process Management}
\label{subsec:llmAndBpm}

This subsection explores the promising potential and ongoing advancements of Large Language Models (LLMs), their application in Business Process Management (BPM), and the challenges that need to be addressed to effectively utilize these models in the BPM context.

Large language models (LLMs), such as GPT-3, have shown remarkable capabilities in a broad array of natural language processing (NLP) tasks, extending their usefulness beyond text generation into more specialized fields, including business process management (BPM). As detailed in \cite{DBLP:conf/bpmds/BuschRSL23}, these LLMs have been effectively utilized for predictive process monitoring and process extraction from text. However, the application of such models typically demands fine-tuning, a process that requires an extensive amount of suitable training data. One innovative approach to bypass this necessity is prompt engineering, which exploits the pre-training of LLMs without the need for additional fine-tuning. This paper outlines a research agenda for incorporating prompt engineering into BPM, thereby broadening the capabilities of LLMs within this discipline.

Moreover, the study presented in \cite{DBLP:journals/corr/abs-2304-04309} emphasizes the lack of systematic investigation into the opportunities offered by LLMs in BPM. Notwithstanding the impressive progress LLMs have made on various tasks, their application to the BPM lifecycle remains less explored. This research foregrounds several BPM tasks and highlights six research directions that call for attention when using LLMs, providing usage guidelines for practitioners.

In \cite{DBLP:journals/corr/abs-2304-11065}, the authors address the rising interest in the use of chatbots, like ChatGPT, for Business Process Management (BPM) applications. The authors aim to systematically assess existing chatbots for their capacity to support conversational process modelling, a critical function in process-oriented capabilities.
They commence by identifying application scenarios across the process life cycle and then conduct a systematic literature review on conversational process modelling. The outcome of this review is a taxonomy that serves as an input for pinpointing application scenarios for conversational process modelling. These scenarios include the paraphrasing and improvement of process descriptions.
The authors evaluate these application scenarios for existing chatbots using a real-world test set derived from the higher education domain. This test set comprises process descriptions, their corresponding process models, and an evaluation of the model quality. The authors leverage the insights from the literature review and the scenario analyses to provide practical implications for the usage and potential avenues for further development of conversational process modelling.

In our previous work \cite{DBLP:journals/corr/abs-2307-02194}, we performed a first exploration of the usage of Large Language Models (LLMs) in the domain of process mining. A key challenge we addressed was the size constraints that limited the amount of data we could feed into the LLM.
To overcome this, we first proposed a methodology to abstract information from standard process mining artifacts. This abstraction allowed us to distill complex data into a more concise format that was manageable for the LLM. Secondly, we developed strategic prompts that guided the LLMs to generate insightful and relevant responses from the abstracted inputs.
However, we did not cover more advanced abstractions (machine learning features, object-centric event logs/process models, \ldots) or multi-prompt strategies.
Also, the evaluation was limited to the GPT-4 large language model.

\iffalse
\section{Discussion}
\label{sec:discussion}

\section{Conclusion}
\label{sec:conclusion}

\section*{Declarations}

{\bf Ethical Approval}: Not Applicable.

\noindent
{\bf Availability of supporting data}: Not Applicable.

\noindent
{\bf Competing Interests}: The authors declare that they have no known competing financial interests or personal relationships that could have appeared to influence the work reported in this paper.

\noindent
{\bf Funding}: Alexander von Humboldt (AvH) Stiftung.

\noindent
{\bf Authors' contributions}: In the execution of this study, Alessandro Berti was primarily responsible for the conceptualization and implementation of the experimental work. Mahnaz Sadat Qafari contributed by detailing and validating the concepts related to fairness that were utilized in the research. Additionally, Daniel Schuster provided valuable insights that significantly enhanced the overall quality of the paper.

\noindent
{\bf Acknowledgements}: We thank the Alexander von Humboldt (AvH) Stiftung for supporting our research.
\fi

\newpage
\clearpage

\bibliography{references}% common bib file

\end{document}